\documentclass[10pt, journal]{IEEEtran}
\pdfoutput=1
\IEEEoverridecommandlockouts
\usepackage{cite}
\usepackage{amsmath,amssymb,amsfonts}
\usepackage{algorithmic}
\usepackage{graphicx}
\usepackage{textcomp}
\usepackage{xcolor}
\def\BibTeX{{\rm B\kern-.05em{\sc i\kern-.025em b}\kern-.08em
    T\kern-.1667em\lower.7ex\hbox{E}\kern-.125emX}}

\IEEEpubid{\makebox[\columnwidth]{978-1-6654-3274-0/21/\$31.00~\copyright~2021~IEEE \hfill}
\hspace{\columnsep}\makebox[\columnwidth]{ }}

\usepackage{pifont}
\usepackage{framed}
\usepackage{soul}
\usepackage{booktabs}  
\usepackage{csquotes}
\usepackage{multirow}
\usepackage{array}
\usepackage{acronym}
\usepackage[linesnumbered,ruled]{algorithm2e}
\newcolumntype{L}[1]{>{\raggedright\let\newline\\\arraybackslash\hspace{0pt}}m{#1}}
\newcolumntype{C}[1]{>{\centering\let\newline\\\arraybackslash\hspace{0pt}}m{#1}}
\newcolumntype{R}[1]{>{\raggedleft\let\newline\\\arraybackslash\hspace{0pt}}m{#1}}
\usepackage{url}

\usepackage{balance}
\usepackage{flushend}

\MakeOuterQuote{"}
\usepackage[caption=false,font=footnotesize]{subfig} % caption=false is needed for IEEE template

\newcommand{\RNum}[1]{\uppercase\expandafter{\romannumeral #1\relax}}

\newcommand{\ie}{\mbox{\emph{i.e.\ }}}

\newcommand{\etal}{\mbox{\emph{et al.\ }}}
\def\eg{{\it e.g. }}
\begin{document}

%\title{Half-Lock Trojans: Reducing Logic Locking \\to Hardware Trojans}
% \title{Logic Locking based Trojans: A Friend Turns Foe}
\title{TroLL: Exploiting Structural Similarities between Logic Locking and Hardware Trojans}

\author{\IEEEauthorblockN{ Yuntao Liu\textsuperscript{1} , \IEEEmembership{Member,~IEEE}, Aruna Jayasena\textsuperscript{2} , \IEEEmembership{Member,~IEEE}, \\ Prabhat Mishra\textsuperscript{3} , \IEEEmembership{Fellow,~IEEE}, Ankur Srivastava\textsuperscript{4} , \IEEEmembership{Fellow,~IEEE}}
\\ \IEEEauthorblockA{\textsuperscript{1}\textit{Department of Electrical and Computer Engineering, Lehigh University} \\
\IEEEauthorblockA{\textsuperscript{2}\textit{Department of Computer Science and Engineering, University of Tennessee at Chattanooga} \\
\IEEEauthorblockA{\textsuperscript{3}\textit{Department of Electrical and Computer Engineering, University of Florida} \\
\textsuperscript{4}\textit{Department of Electrical and Computer Engineering \& Institute for Systems Research,\\  University of Maryland, College Park} \\
\textsuperscript{1}yule24@lehigh.edu, \textsuperscript{2}aruna@tennessee.edu, \textsuperscript{3}prabhat@ufl.edu, \textsuperscript{4}ankurs@umd.edu }}}}

% \and
% 

\maketitle
\begin{abstract}
Logic locking and hardware Trojans are two fields in hardware security that have been mostly developed independently from each other. In this paper, we identify the relationship between these two fields. We find that a common structure that exists in many logic locking techniques has desirable properties of hardware Trojans (HWT). We then construct a novel type of HWT, called Trojans based on Logic Locking (TroLL), in a way that can evade state-of-the-art ATPG-based HWT detection techniques. 
In an effort to detect TroLL, we propose customization of existing state-of-the-art ATPG-based HWT detection approaches as well as adapting the SAT-based attacks on logic locking to HWT detection.
% Experiments show that TroLL is extremely resilient to both existing ATPG-based HWT detection techniques and SAT-based detection: the percentage of TroLL detected falls drastically and the trigger length increases. In fact, none of these detection approaches could detect more than 1\% of TroLL with 24-bit triggers.
In our experiments, we use random sampling as reference. It is shown that the customized ATPG-based approaches are the best performing but only offer limited improvement over random sampling. Moreover, their efficacy also diminishes as TroLL's triggers become longer (\textit{i. e.} have more bits specified).
We thereby highlight the need to find a scalable HWT detection approach for TroLL.

\end{abstract}


%   <concept>
%       <concept_id>10010583.10010633</concept_id>
%       <concept_desc>Hardware~Very large scale integration design</concept_desc>
%       <concept_significance>500</concept_significance>
%       </concept>
%   <concept>
%       <concept_id>10010583.10010682.10010712</concept_id>
%       <concept_desc>Hardware~Methodologies for EDA</concept_desc>
%       <concept_significance>500</concept_significance>
%       </concept>
%   <concept>
%       <concept_id>10002978.10003001</concept_id>
%       <concept_desc>Security and privacy~Security in hardware</concept_desc>
%       <concept_significance>500</concept_significance>
%       </concept>
%  </ccs2012>

% \ccsdesc[500]{Hardware~Very large scale integration design}
% \ccsdesc[500]{Hardware~Methodologies for EDA}
% \ccsdesc[500]{Security and privacy~Security in hardware}

% \keywords{
\section{Introduction}
The fact that most chip designers outsource the production of their chips to off-shore foundries raises concerns about the privacy of the chip's intellectual property (IP) and the integrity of the fabrication process. There has been a significant amount of research in both topics. For IP protection, numerous design obfuscation techniques have been proposed to mitigate attacks such as counterfeiting and over production, among which logic locking is by far the most prominent and well-studied class of protection techniques \cite{chakraborty2019keynote}. Logic locking adds key inputs and key-controlled gates into the circuit to make the locked circuit's functionality key-dependent. As the correct key is not known to the untrusted foundry, neither is the correct functionality, and hence the privacy of the design is preserved.
Pertaining to the integrity of fabrication, the term Hardware Trojans (HWT) is often used to describe stealthy malicious modifications in the design.
Logic locking and HWT's have been studied mostly independently so far. 
% \color{blue}
Existing literature have studied the impact of logic locking and HWT's on each other, such as how to use design obfuscation to prevent HWT insertion \cite{frey2015exploiting,hu2019leveraging,yu2017exploiting,dupuis2014novel,mirmohammadi2023new,cruz2022analysis,wang2024trolloc} and inserting HWT's in logic locked circuits \cite{chakraborty2009security,maynard2024reconfigurable}. These works treated logic locking and HWT's as two separate subjects and do not study their similarity.
In contrast, in this work, we uncover the structural similarity between logic locking and HWT's. Leveraging this insight,
% , little attention was paid to the relationship between logic locking and HWT’s. 
% Apart from some studies on how to use design obfuscation to prevent HWT insertion \cite{frey2015exploiting,hu2019leveraging,yu2017exploiting,dupuis2014novel,mirmohammadi2023new,cruz2022analysis,maynard2024reconfigurable} and how to compromise obfuscation with HWT’s \cite{chakraborty2009security}, little attention was paid to the relationship between logic locking and HWT’s. 
%\color{black}
we will discuss how to utilize logic locking techniques to construct novel HWT's. 
%\color{blue}
This is a significant threat because logic locking techniques have become public knowledge and adversaries can leverage the principles and structures of logic locking to craft novel HWT's. However, this threat has been largely overlooked in the current security landscape. Our research highlights how adversaries can exploit logic locking mechanisms to create hard-to-detect Trojans. We also explore how to convert attacks against design obfuscation to HWT detection techniques. We demonstrate that such Trojans are resilient not only to existing detection techniques but also to newly customized methods tailored to detect such Trojans.
%\color{black}
The contribution of this work is as follows.

%\color{blue}
\begin{itemize}
    \item We present a novel perspective on logic locking by identifying structural similarities between logic locking schemes and hardware Trojans (HWT's). In particular, we decompose logic locking mechanisms into a \textit{Mutation Unit (MU)} and a \textit{Restore Unit (RU)}, laying the groundwork for cross-domain insights.
    
    \item Building on this insight, we propose a new class of Trojans, \underline{Tro}jans based on \underline{L}ogic \underline{L}ocking (TroLL), by embedding only the MU into the circuit and bypassing the RU. This departs from traditional HWT design strategies that rely on rare-event triggers.
    
    \item We adapt state-of-the-art Automatic Test Pattern Generation (ATPG) techniques to account for TroLL’s trigger characteristics and assess their effectiveness against this new class of Trojans.

    \item We further extend logic locking attack strategies, specifically SAT-based attacks, to evaluate their applicability in detecting both TroLL and conventional HWTs.

    \item Our experimental results reveal that TroLL exhibits strong resilience to existing ATPG-based detection techniques such as statistical test generation~\cite{chakraborty2009mero} and maximum clique sampling~\cite{lyu2020scalable}. While adapted detection methods perform better on TroLL without sacrificing effectiveness on conventional HWTs, their performance degrades significantly as the trigger length increases.
\end{itemize}

The rest of this paper is organized as follows. In Section \ref{sec:bg}, we introduce the technical background of hardware Trojans and logic locking. Section \ref{sec:troll} presents the structural similarities between logic locking techniques and hardware Trojans, and the construction of TroLL based on such similarities. The evolved ATPG-based detection approaches and the adaptation of SAT-based attacks on logic locking to HWT detection is formulated in Section \ref{sec:troll_detect}. In Section \ref{sec:exp}, we present the experiment details on TroLL and the results on detecting TroLL with approaches based on ATPG, SAT, and random sampling. Lastly, we conclude the paper in Section \ref{sec:conc}.

% \vspace{-3mm}
\section{Background and Related Work} \label{sec:bg}
% \vspace{-1mm}

In this section, we provide relevant background and survey related efforts in three broad categories. First, we describe the working principle of hardware Trojans. Next, we survey existing test generation efforts for detection of hardware Trojans. Finally, we provide an overview of logic locking techniques.

\subsection{Hardware Trojans} \label{ssec:hwt_intro}
% \vspace{-1mm}

\begin{figure}
    \centering
    \includegraphics[width=.48\textwidth]{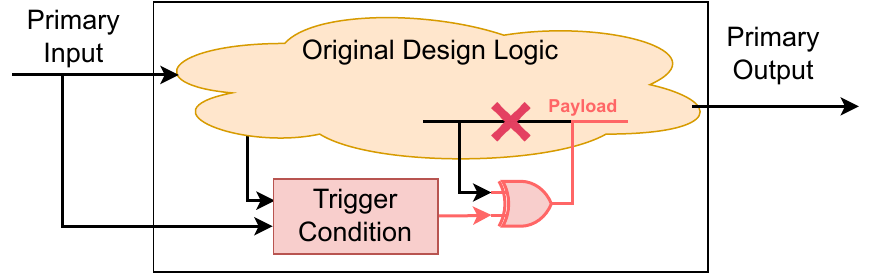}
    \caption{Illustration of an HWT-infested Circuit}
    \vspace{-5mm}
    \label{fig:trojan_example}
\end{figure}

Hardware Trojans (HWT) are stealthy malicious modifications to hardware designs.
HWT's usually consist of two components, trigger and payload. The trigger is a condition that activates the HWT, and the payload is the effect of the HWT once activated.
The trigger condition can be determined by the circuit's input and/or internal signals in the original design.
The HWT payload can have various possible effects, including functionality corruption \cite{zhang2013hardware}, information leakage \cite{tsoutsos2014advanced, fern2016hiding, jin2009experiences}, denial-of-service \cite{reece2012stealth}, bypass of security properties \cite{sturton2011defeating}, etc. An illustration of an HWT-infested circuit is given in Fig. \ref{fig:trojan_example} where the relationship between the original design and the HWT's trigger and payload is shown.

HWT's can be inserted in almost any phase in the VLSI hardware supply chain, including in third-party IP cores, by a malicious CAD tool, by a rogue designer, by an untrusted fabrication facility, etc. \cite{xue2020ten, karri2010trustworthy}.
The HWT's inserted before the fabrication stage are present in the design files (\eg RTL and/or gate-level netlists). Therefore, it is possible to use formal methods, such as logic equivalence checking, to tell whether an HWT exists \cite{rathmair2014applied, fern2017detecting}. However, for HWT's inserted by the foundry, the netlist of the HWT-infested circuit is not available to the designer. Some researchers have proposed to use reverse engineering to obtain the layout of the HWT-suspicious chip \cite{bao2017reverse, vashistha2021detecting}. However, IC reverse engineering is increasingly expensive and error-prone as technology node scales down \cite{torrance2011state}
% , and there is no report of successful reverse engineering of any chip with technology node below 14nm to the best of our knowledge. 
 To date, the smallest technology node that is publicly known to be successfully reverse engineered down to the netlist at the whole chip scale is 14nm \cite{waite2021preparation}, and the smallest node demonstrated for reverse engineering-based HWT detection is 20nm \cite{krachenfels2023trojan}. 
The netlist in state-of-the-art (SOTA) technology node is at 3nm \cite{TSMC_Q2_2025_transcript, Samsung_3nm_press_2022}, which means the reverse engineering-based HWT detection toolchain is a few generations behind, making it infeasible to perform netlist-based HWT detection for chips produced at the latest nodes.
Hence, testing is the most practical way to detect HWT's inserted by untrusted foundries. Besides, testing-based methoes are also applicable to HWT's inserted by IP providers, CAD tools, rogue designers, etc. The state-of-the-art automatic test pattern generation (ATPG) approaches for HWT detection will be introduced in Section \ref{ssec:atpg}.

% The signal rarity analysis can be done at either the register transfer (RT) level \cite{salmani2013analyzing} or the gate level \cite{cruz2018automated, abbassi2019trojanzero}. 
%Using a combination of rare signals as a trigger ensures that the Trojan-infested circuit still behaves correctly under most circumstances and hence the Trojan is not easily detected.

\subsection{ATPG-based HWT Detection} \label{ssec:atpg}
Both combinational and sequential HWT triggering mechanisms have been proposed in the literature. However, since the designer likely has access to testing infrastructure that allows the circuit to be tested combinationally (\eg scan-chain), a sequential HWT trigger can be broken down into a series of combinational ones. We hence focus on combinational HWT triggers in this work.
State-of-the-art combinational HWT insertion methodology utilizes rare signals (\ie an internal node's value that is functionally difficult to sensitize) as the trigger, ensuring that the HWT is only triggered in rare circumstances \cite{cruz2018automated, abbassi2019trojanzero}. Based on this property, many HWT detection methods have been developed based on ATPG principles.
% which we will introduce in Section \ref{ssec:atpg}.
% ATPG-based HWT detection approaches also assume that HWT's are triggered by internal rare signals.
% an adversary is likely to construct a trigger using rare signals such that the HWT stays hidden during functional validation using millions of test patterns. 
% Hence, these approaches utilize principles of ATPG to sensitize the rare signals.
% The goal of the existing approaches is to sensitize the rare signals as much as possible. 
Existing approaches explored two complementary directions when dealing with test generation for activation of rare signals: 1) statistical test generation, and 2) directed test generation. 
A promising avenue for statistical test generation is to rely on $N$-detect principle~\cite{pomeranz2004measure} by activating each rare signal $N$ times to increase the statistical likelihood of activating the unknown trigger in the HWT. MERO~\cite{chakraborty2009mero} tries to generate test vectors to activate the same rare signal $N$ times by flipping input vector bits one at a time. Saha \etal improved the test generation performance using genetic algorithm and Boolean satisfiability~\cite{saha2015improved}. Pan \etal improved the performance further by flipping bits using reinforcement learning~\cite{pan2021automated}. 

While $N$-detect methods try to activate one rare signal at a time, Lyu \etal focused on activating as many rare nodes as possible using maximal clique sampling (TARMAC~\cite{lyu2020scalable}). TARMAC first creates the satisfiability graph for all the rare signals. {
In this graph, each vertex stands for a rare signal, and there is an edge connecting two vertices if and only if there exists an input pattern that sensitizes the two rare signals simultaneously.}
Next, the maximal cliques from the satisfiability graph is computed. Finally, TARMAC generates tests to activate randomly sampled set of maximal cliques. If any of the generated tests is able to activate the trigger, the HWT will be detected.

% \vspace{-2mm}
\subsection{Logic Locking} \label{ssec:locking}
% \vspace{-1mm}

Logic locking has emerged as a protection mechanism against potential piracy and overbuilding threats in untrusted fabrication facilities. These techniques obfuscates the hardware by adding key inputs into the circuit without disclosing the correct key to the fab. Hence, the fab will not know the full functionality of the design. When the fabrication is done, the chip designer (or a trusted third party) will provide the key to the chip by connecting a piece of tamper proof memory. This process is called \textit{activation}. This way, only the authorized users will have access to an \textit{activated chip} which has the correct functionality. 

\begin{figure}
    \centering
    \includegraphics[width=.48\textwidth]{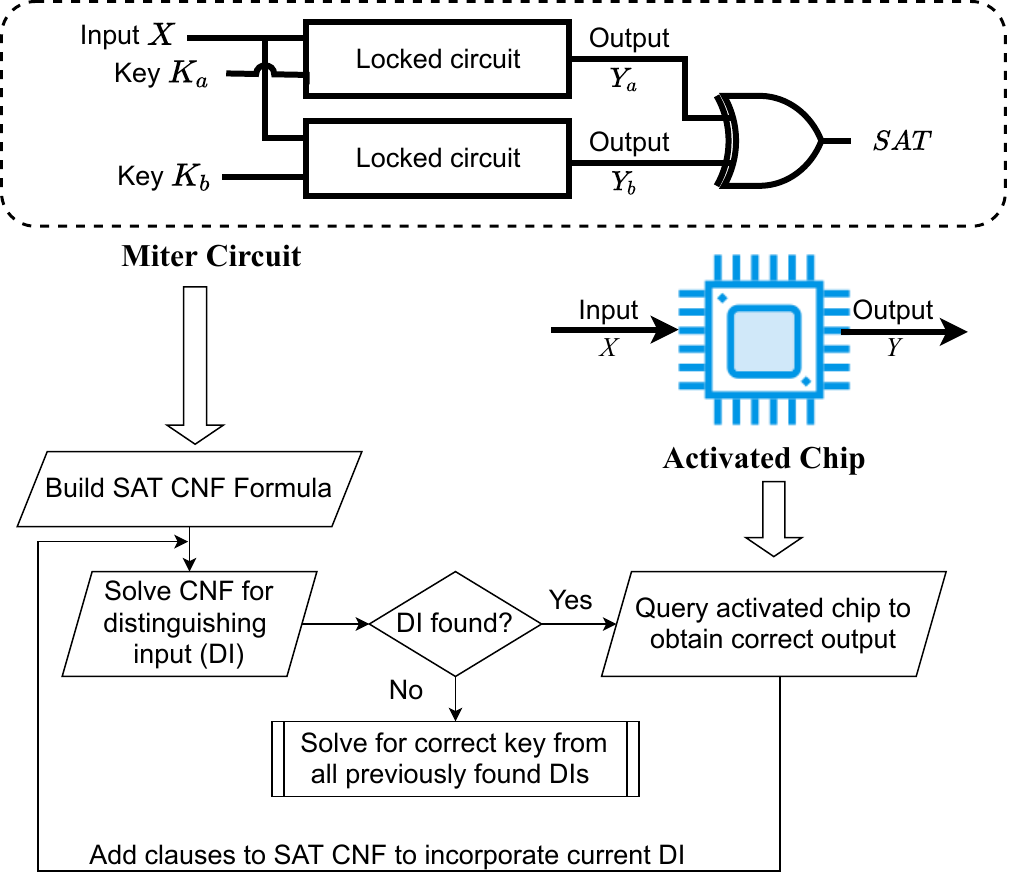}
    \vspace{-3mm}
    \caption{The basic procedure of SAT-based attacks}
    \vspace{-5mm}
    \label{fig:sat_attack}
\end{figure}

There have been many attacks formulated against logic locking, among which the ones based on Boolean satisfiability theory, a.k.a. SAT-based attacks \cite{subramanyan2015evaluating}, have both mathematical guarantee to find the correct key and strong practical performance. 
The flow of SAT-based attacks is demonstrated in Fig.~\ref{fig:sat_attack}. As demonstrated, a miter circuit is built. The miter contains two copies of the locked netlist that share the same input but are keyed separately. Their outputs are XOR'ed. Essentially, if the miter's output is $TRUE$, the input is causing different outputs with the two keys. The SAT-based attacks are iterative. In each iteration, a Boolean satisfiability problem is solved to find an input pattern and two keys that satisfy the miter circuit. The input pattern is called the \textit{distinguishing input (DI)}. The activated chip is then queried to get the correct output value. Then, clauses are added to the miter-based SAT formula so that all the wrong keys that causes an incorrect output for the DI are pruned out. A correct key will be found when the DIs found have pruned out all the wrong keys.

Many SAT resilient logic locking techniques have been proposed to thwart the attack. In this work, we will examine these techniques and summarize the structural similarities among them. We then show how these logic locking techniques can guide the construction of novel hardware Trojans.

% \vspace{-3mm}
\section{Locking Inspired HWT Construction} \label{sec:troll}
% \vspace{-1mm}
In this section, we provide a brief overview of existing obfuscation art. We then explore how the properties of these techniques can be leveraged in order to construct difficult-to-detect HWT's by slightly modifying their logical topologies, maintaining their rigorous mathematical guarantees but retargeting them to HWT application.
% In this section, we describe how to convert a logic locking technique to a HWT insertion algorithm. 
The intuition behind such conversion is that, for both locking and Trojan, error is injected into a circuit only when the circuit has a specific input pattern:
\begin{itemize}
    \item For locking: The input pattern is among those that are corrupted by the given wrong key.
    \item For Trojans: The input pattern matches the trigger.
\end{itemize}
% effect of HWT's can be realized by the output corruption of a locked circuit due to a wrong key. 
Because HWT's should be triggered only by very few input patterns to evade detection \cite{chakraborty2009mero, lyu2020scalable}, the logic locking schemes suitable for converting to HWT's should also corrupt very few input patterns given a wrong key. 
Such logic locking techniques do exist and they are mainly designed to thwart SAT-based attacks. These techniques include Anti-SAT \cite{xie2016mitigating}, SARLock \cite{yasin2016sarlock}, stripped functionality logic locking (SFLL) \cite{yasin2017provably}, Robust Strong Anti-SAT \cite{liu2021robust}, CASLock \cite{shakya2020cas}, etc. In this work, we first analyze the commonality among these locking approaches. Next, we present the HWT construction based on these locking algorithms. 

\begin{figure*}[t]
\centering
\subfloat[SFLL\label{fig:sfll-mu}]{
    \centering
    \includegraphics[width=.31\textwidth]{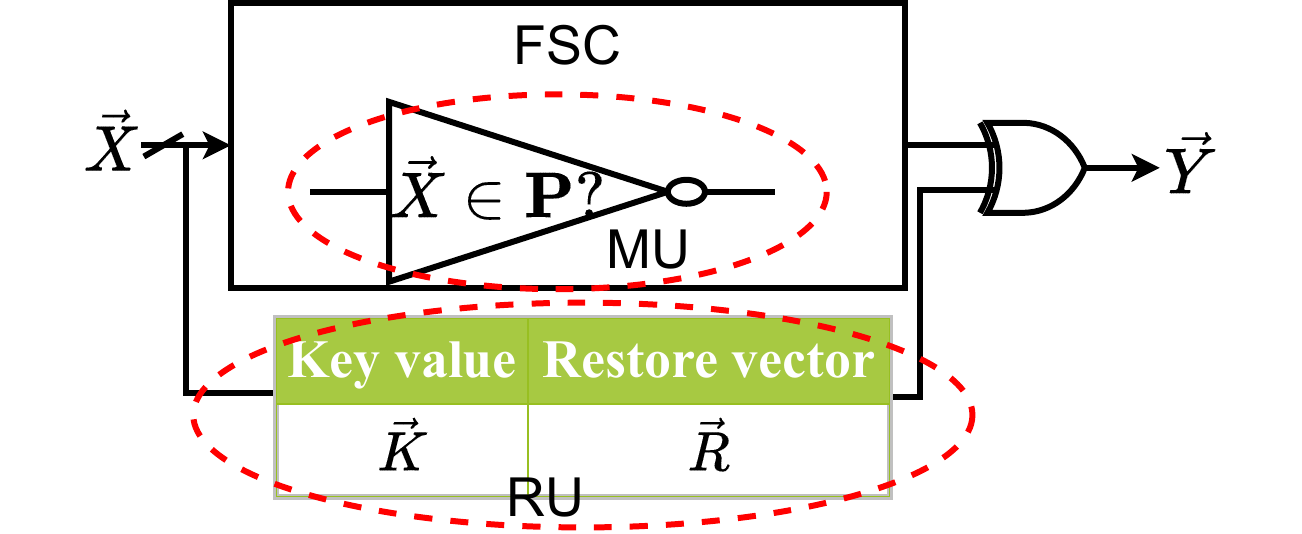}
    }
    \hfill
\subfloat[Anti-SAT\label{fig:anti-sat-mu}]{
    \centering
    \includegraphics[width=.31\textwidth]{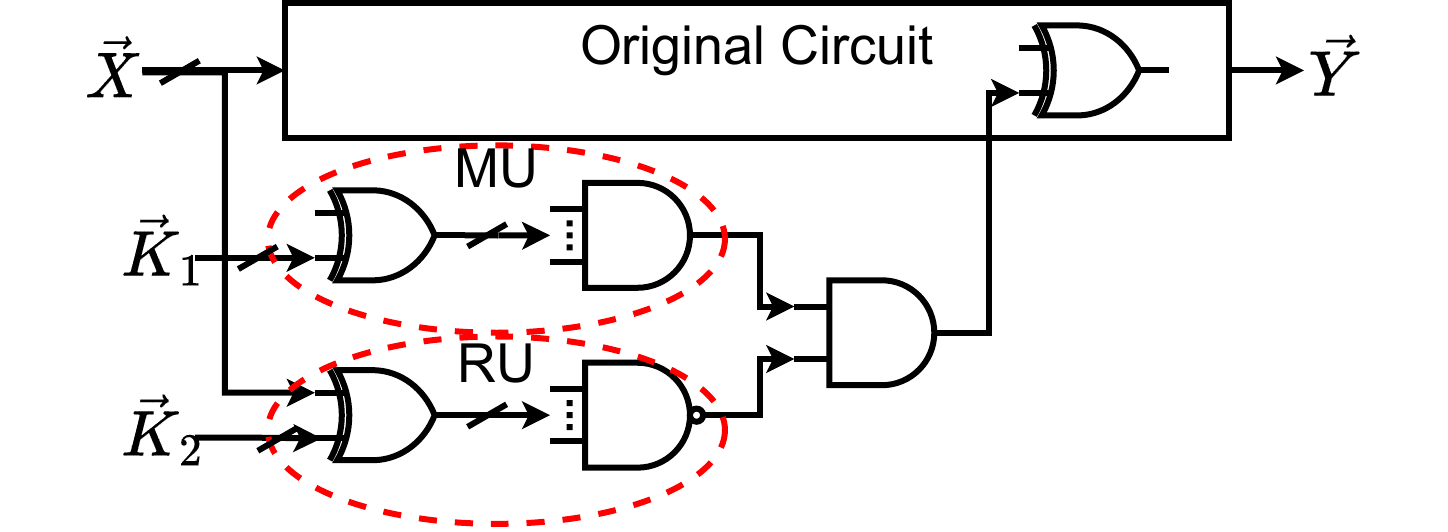}
    }
    \hfill
\subfloat[SARLock\label{fig:sarlock-mu}]{
    \centering
    \includegraphics[width=.31\textwidth, trim={0 1.5cm 0 0}, clip]{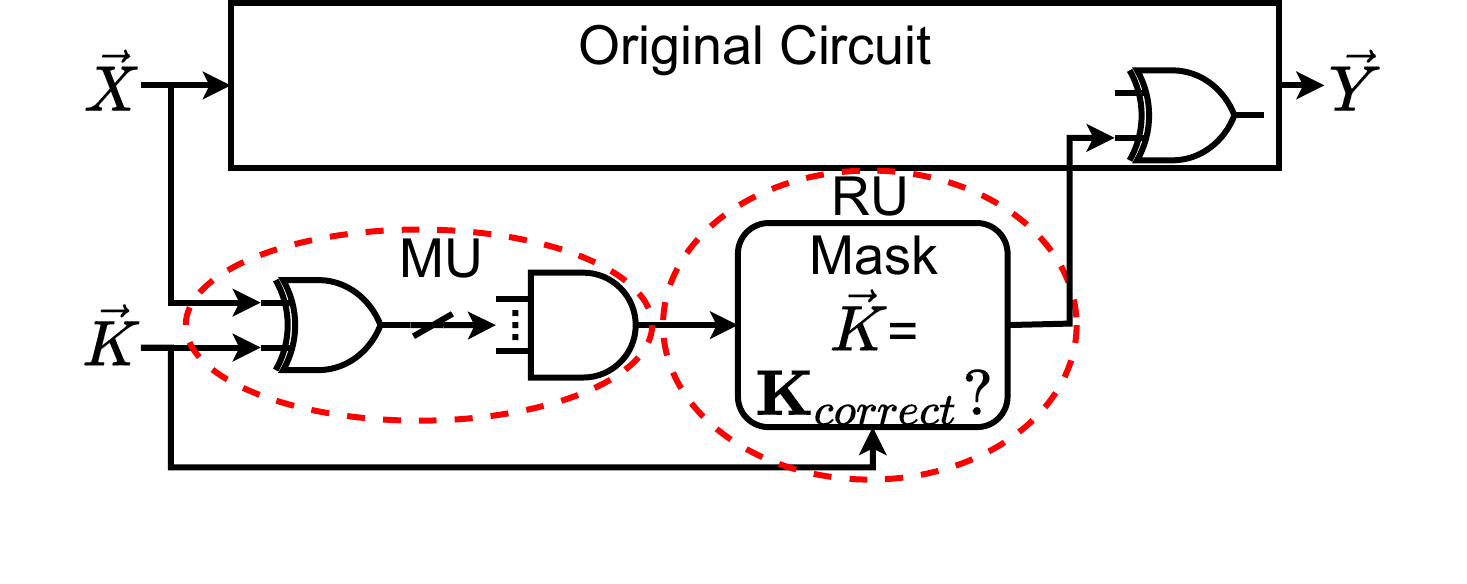}
    }
% \vspace{-5mm}
\caption{MU and RU in Logic Locking Constructions}
\vspace{-3mm}
\end{figure*}

% \vspace{-3mm}
\subsection{Commonality among Logic Locking}
% \vspace{-1mm}
% Although the above-mentioned SAT-resilient locking techniques have distinctly different constructions, 
No matter how distinct these logic locking constructions seem to be, we find that they can all be decomposed into two functional units that interact together to inject error for specific input pattern given a wrong key. We call them a \textit{Mutation Unit (MU)} and a \textit{Restore Unit (RU)}. 
Essentially, the MU modifies the circuit's functionality for some input patterns and the RU tries to restore the modified functionality.
When the correct key is applied, the RU restores the correct input patterns modified by the MU and so the locked circuit produces correct output to all input values. When the key is incorrect, however, the error injected by the MUs will not be corrected by the RU. 
In this case, if the input's functionality is modified by the MU, its output will be corrupted. The number of input patterns modified by the MU should be very small in order for the logic locking approach to be resistant to SAT-based attacks~\cite{liu2020strong}.
The rarity of such input patterns makes them suitable for HWT triggers.
% According to the unavoidable trade-off between the error rate and expected number of SAT attack iterations for any logic locking technique \cite{liu2020strong}, the number of such input patterns must be very small compared to the size of the input space which makes these locking techniques resilient to SAT attacks. 
% It is worth noting that there are many forms that MU's and RU's can be in: conditional inverters, AND-trees, look-up tables (LUT), etc. 
We use SFLL, SARLock, and Anti-SAT as examples of SAT-resilient locking techniques and briefly review how the MU and RU interact in each of them. 
%Let us briefly overview each of them to begin.

% \vspace{-2mm}
\subsubsection{Stripped Functionality Logic Locking (SFLL)}
Fig. \ref{fig:sfll-mu} shows the block diagram of SFLL. It is composed of two parts: a functionality stripped circuit (FSC) and a restore unit (RU). The FSC is the original circuit with the functionality altered for a set of protected input patterns (PIP), denoted as $\mathbf{P}$. The FSC's internal structure that modifies the functionality of the PIPs is the MU of SFLL. Notice that RU of SFLL coincides with our general definition of RU. The structure of the RU in SFLL is a look-up table (LUT). If the circuit's input matches the LUT key, the LUT will produce a restore signal. If the LUT contains the correct key, the restore signal will reverse the corruption caused by the FSC. If the LUT contains an incorrect key, both the PIPs and the input patterns that correspond to the key will be corrupted.

% \vspace{-2mm}
\subsubsection{Anti-SAT}
The structure of Anti-SAT is shown in Fig. \ref{fig:anti-sat-mu}. The MU and the RU have similar structure. For the MU, there is an array of XOR gates followed by an AND tree. Depending on $\vec{K}_1$, there is only one input value of $\vec{X}$ such that the MU will evaluate to logic 1. Let us call this value $\vec{X}_M$. The RU's structure is very similar to the MU's, and the only difference is that the AND tree's output is inverted. Depending on $\vec{K}_2$, there is only one input value of $\vec{X}$ that will make the RU evaluate to logic 0. Let us call this value $\vec{X}_R$. Corruption is injected into the circuit when both the MU and the RU evaluate to logic 1, \ie when $\vec{X}=\vec{X}_M$ and $\vec{X}\ne\vec{X}_R$. Therefore, a correct key must be such that $\vec{X}_M=\vec{X}_R$. This way, the RU will output logic 0 when MU outputs logic 1 and prevent the original circuit from being corrupted.

% \vspace{-2mm}
\subsubsection{SARLock}

SARLock also contains an MU and an RU, as shown in Fig. \ref{fig:sarlock-mu}. Its MU is the same as the one in Anti-SAT: depending on the key, there is one input value that will let the MU evaluate to logic 1. The RU checks if the key input contains the correct key. If  so, it will mask the MU's output and prevent it from corrupting the original circuit.

% \vspace{-2mm}
\subsection{Advantages of Locking based Trojans}
% \section{Locking-based HWT Construction}
% \vspace{-1mm}
In each of the above-mentioned logic locking techniques, the MU is capable of injecting error into the circuit, and the RU will prevent the error from being injected if the correct locking key is provided.
We notice that the MU naturally offers properties desirable for the trigger of HWT's: 
\begin{itemize}
    \item Corruption is injected for very few (or just one) input pattern in the exponential input space, which makes random sampling based detection very difficult.
    \item The corrupted input patterns need not have any correlation with the original netlist's structure, so that they can be chosen to avoid ATPG or rare signal based detection approaches such as \cite{chakraborty2009mero} and \cite{lyu2020scalable}.
    % satisfying the rareness requirement of HWT triggers.
    \item These trigger patterns are completely known to the attacker. Contrarily, enumerating triggers of conventional rare signal based Trojans is mathematically intractable in general because it is a satisfiability counting problem \cite{creignou1996complexity}. Hence, it is much easier for the attacker to control when to trigger the Trojan and avoid unintended triggering using TroLL.
    % satisfying the controllability requirement of HWT triggers.
\end{itemize}

% \vspace{-3mm}
\subsection{Construction of TroLL}
% \vspace{-2mm}
These properties indicate that the MU's of logic locking can serve as ideal HWT trigger circuitry. Building upon this discovery, we present Trojans based on logic locking (TroLL), which employs the MU of logic locking to modify the functionality of the original circuit. We present a generalizable way to convert a logic locking technique to TroLL as follows:
\begin{enumerate}
    \item Identify MU and RU in the locked netlist and remove the RU. Hard-code the RU's output value to the one that does not invert the output of the MU.
    \item If the MU has a key input (such as Anti-SAT), hard-code the key such that the desired HWT trigger can cause the MU to corrupt the circuit.
\end{enumerate}
Essentially, when building TroLL from SFLL, we only need to remove the RU and make sure that the PIP's represent the Trojan trigger patterns we want. For Anti-SAT and SARLock, we need to remove the RU and hard-code the MU keys to incorporate the triggers. E.g., for the Anti-SAT construction in Fig. \ref{fig:anti-sat-mu}, we need to remove the RU and fix its output at logic 1. For the MU, we fix $\vec{K}_1$ to be the bitwise-inverted trigger pattern.
A constant sweep is then performed to simplify the circuit. In this way, the key inputs of logic locking will be all removed and the TroLL-infested circuit has the same I/O pins as the original circuit. No matter which logic locking technique TroLL is made from, the functionality of TroLL will be identical. Besides, as each of the above steps is a strict subtraction from logic locking infrastructure, TroLL's overhead will be much lower than that of logic locking. Notice that, although we describe a gate-level operation to build TroLL in the above example, TroLL can be incorporated at RT or behavioral level using the two-step process as well.

% \vspace{-3mm}
\subsection{Choosing TroLL Trigger Pattern} \label{ssec:trigger_select}
% \vspace{-1mm}
\begin{algorithm}[htb]
\SetKwInOut{Input}{input}
\SetKwInOut{Output}{output}
\Input{$G_1\ldots G_n, r_1\ldots r_n, p_1\ldots p_n$; \tcp{\it The Boolean function, rare output value and associated probability of each gate in the original design}}
\Input{$\mathcal{S}$\tcp*{\it A set of random input vectors}}
\Output{$X_T$ \tcp*{\it The best TroLL trigger found}}
\Output{$p_{max}$ \tcp*{\it Maximum rare value probability threshold}}
\BlankLine\emph{Initialization: $p_{max}\gets 0$\; \label{line:initialize}} 
\ForEach{$X\in\mathcal{S}$\label{line:for_input}}{
    $p_{tmp} \gets 0.5$ \tcp*{\it tracking minimum rare probability for each sample}
    \ForEach{$i \in \{1\ldots n\}$\label{line:for_gate}}{
        \tcp{\it Check if gate $i$ has the rare value}
        \If{$G_i(X)==r_i$}{
          \If{$p_{tmp}>p_i$}{
                $p_{tmp}\gets p_i$\;
            }
        }
    }
    \If{$p_{tmp}>p_{max}$}{
        $p_{max}\gets p_{tmp}$\;
        $X_T \gets X$\;
    }
}
\caption{TroLL Trigger Selection}
\label{algo:trigger}
\end{algorithm}

\begin{figure}[tb]
    \centering
    \includegraphics[width=.48\textwidth]{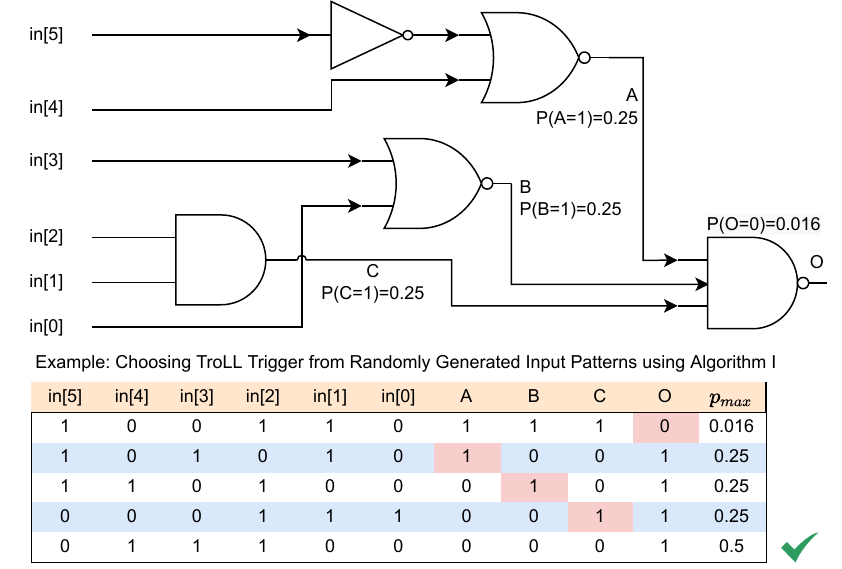}
    \caption{Illustration of how to Choose TroLL Trigger using Algorithm \ref{algo:trigger} on a Sample Circuit}
    \vspace{-5mm}
    \label{fig:algo_ex}
\end{figure}

TroLL needs to evade HWT detection. As introduced in Section \ref{ssec:hwt_intro}, existing state-of-the-art HWT detection approaches find test patterns that sensitize rare signals in the original design. To evade these detection approaches, TroLL trigger patterns need to avoid sensitizing any rare nodes. 
To begin with, we use a random sampling approach to determine the rare value of each internal node, $r_i$, and its associated probability, $p_i$. Although an alternative to random sampling is the signal probability skew analysis \cite{yasin2017security}, the complexity of such analysis often increases exponentially if the correlation between signals is to be accounted for \cite{mohyuddin2011probabilistic}.
Then we use Algorithm \ref{algo:trigger} to determine the trigger pattern for TroLL. Essentially, the algorithm finds an input pattern with the maximum probability threshold $p_{max}$ such that no rare value below this probability will be realized by the trigger. 
Such a process is illustrated in Fig. \ref{fig:algo_ex}. In the sample circuit, the rare values and their probabilities are annotated for each internal node. A list of randomly generated input patterns are shown under the circuit diagram. The signal sensitized by each input pattern that has the lowest probability are highlighted in pink. Algorithm \ref{algo:trigger} will choose the input pattern that maximizes the lowest probability. In this example, the trigger pattern will be the one in the last row since it does not sensitize any rare value.
TroLL triggers selected by this process will be immune to the existing rare value based detection approaches such as those introduced in Section \ref{ssec:hwt_intro}. 

The fact that TroLL triggers does not sensitize any rare signal does not mean that TroLL can be triggered by high probability signals or can be easily detect by random sampling. 
On the contrary, TroLL is essentially creating a new rare node that only the trigger pattern can sensitize. Since the defender does not have the netlist of the HWT-infested circuit and can only base the detection on the original circuit, they do not have any information about the new node and hence cannot generate test patterns aimed at sensitizing the new node. 
Also notice that the triggers selected using Algorithm \ref{algo:trigger} has the full input length. This will likely cause high overheads. As we later demonstrate in Sections \ref{ssec:detect_rare}, practical resilience against HWT detection can be attained when only a subset of input bits are taken as the TroLL trigger.

% \subsection{Payload Selection} \label{ssec:payload_select}
% The most common efficacy of HWT payload is to corrupt the circuit's functionality. In this work, we evaluate each internal net's fitness as the payload by calculating the average hamming distance (HD), \ie the number of output bits flipped when the internal net is flipped. All the internal nets are then sorted by the average HD in descending order. We then add the nets from top of the list to the payload until an HD of approximately 50\% is observed at the output when the Trojan is triggered.

\section{Detection Techniques for TroLL} \label{sec:troll_detect}
In this section, we introduce a few novel approaches that are aimed to detect TroLL more effectively. The first type of approaches are based on the trigger selection process of TroLL: by avoiding any test pattern that sensitize any rare node value, ATPG-based HWT detection mechanisms will generate test patterns that are more likely to match the trigger of TroLL. The second approach is based on the fact that TroLL originates from logic locking, and SAT-based attacks are the most formidable attacks on logic locking. Therefore, we can formulate a Trojan detection approach that emulates the a SAT-based attack on logic locking.

\subsection{Customizing ATPG-based HWT Detection Approaches for TroLL} \label{ssec:custom_atpg}
Given TroLL's trigger selection mechanism, we can customize existing ATPG-based HWT detection approaches to detect TroLL. TroLL's trigger selection process eliminates any input pattern that sensitizes any rare internal node value as described in Algorithm \ref{algo:trigger}. The same principles can be applied to the test generation algorithms for HWT detection: instead of targeting the rare values, the ATPG algorithms can choose the test patterns that satisfy as many prevalent values as possible. 
Following the notations used in Algorithm \ref{algo:trigger}: say that $n$ internal nodes of a combinational circuit that implement Boolean functions $G_1\ldots G_n$ have rare values $r_1 \ldots r_n$ that are below a certain threshold $p$ where $0<p<0.5$. In other words, these $n$ nodes have prevalent values $\bar{r_1} \ldots \bar{r_n}$ that have probabilities above $1-p$.
While existing HWT detection algorithms aim to find test patterns $X$ that satisfy as many $G_i(X)=r_i$ as possible ($i=1\ldots n$), a TroLL-specific detection algorithm should instead find input patterns that satisfy $G_i(X)=\bar{r_i}$ for as many $i$ as possible.

Given such a principle, it is surprisingly convenient to customize existing HWT detection approaches for TroLL. We can indeed run the same ATPG algorithms, such as statistical test generation \cite{chakraborty2009mero} or maximal clique sampling \cite{lyu2020scalable}, and target the same set of internal nodes. The only change is to invert the targeted Boolean values of these nodes. Statistical test generation (such as $N$-detect) can target to generate test vectors to activate each prevalent node value $N$ times, whereas maximal clique sampling can build the satisfiability graph on the prevalent values instead of the rare values. 
% In Section \ref{ssec:detect_rare}, we compare the customized versions statistical test generation and maximal clique sampling with their original forms.

Because the defender does not know the type of HWT when the test patterns are generated, the test patterns should be able to detect conventional HWTs as well. Therefore, for each ATPG algorithm, we combine the test patterns that are generated to sensitize the rare values (for conventional Trojans) and those generated to avoid sensitizing rare values (for TroLL). We refer to such an approach as \textit{Evolved Statistical Test Generation} and \textit{Evolved Maximal Clique Sampling}. In Section \ref{ssec:detect_rare}, we will present the efficacy of these evolved HWT detection approaches.

% \vspace{-3mm}
\subsection{Adapting SAT-based Attacks on Logic Locking for HWT Detection} \label{sec:sat}
% \vspace{-1mm}
% SAT-based attack formulations are usually strong against logic obfuscation approaches. 
Attacks on logic locking try to find the correct key, whereas Trojan detection aims to find the trigger of HWT's. Since TroLL is based on logic locking, it is natural to associate logic locking attack with the detection of TroLL. However, since the defender does not know which type of HWT is potentially inserted, the detection approaches must not be limited to TroLL but generalizable to any type of HWT.
% In this section, we present \textit{SAT-based iterative trigger pruning (SITP)}, an HWT detection formulation that is based on the SAT attack against logic locking and iteratively prunes the search space for HWT triggers.
In this section, we present how to adapt the SAT-based attacks on logic locking to detecting HWT's. A SFLL-like auxiliary circuit will be constructed based on the HWT-suspicious circuit where the Trojan's trigger and payload are represented by keys. Then, the SAT attack formulation is used to find a key that can represent the HWT. The HWT is detected when such a key is found. In Section \ref{sec:exp}, this SAT-based detection approach as well as the ATPG-based approaches will be used to evaluate the detectability of TroLL and conventional HWT's.

% \vspace{-3mm}
\subsubsection{Construction of the Auxiliary Circuit}
\begin{figure} [htb]
    \centering
        % \vspace{-0.1in}
    \includegraphics[width=.48\textwidth]{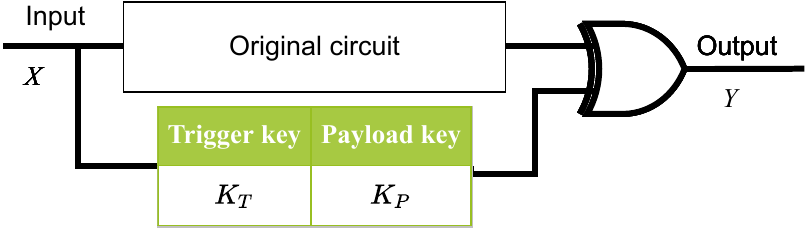}
        % \vspace{-3mm}
    \caption{Construction of the auxiliary circuit for SAT-based detection}
        % \vspace{-1mm}
    \label{fig:sat_aux}
\end{figure}

% \vspace{-1mm}
A defender has the netlist of the original circuit and the fabricated HWT-suspicious circuit. The netlist of the fabricated circuit is not available. In order to search for a trigger pattern, an SFLL-like auxiliary circuit to emulate an HWT-infested circuit is constructed.
As shown in Fig. \ref{fig:sat_aux}, the auxiliary circuit is built by adding a look-up table to emulate the trigger and payload of the HWT. The trigger key $K_T$ is compared with the circuit input $X$. When they are the same, the payload key $K_P$ is bit-wise XOR'ed with the output $Y$.

Note that SAT-based detection does not assume any knowledge information about the potentially existing HWT, and the construction of the auxiliary circuit is independent from the actual trigger and payload of the HWT. The purpose of the auxiliary circuit is to emulate the trigger and payload of HWT's rather than being functionally equivalent to the HWT-suspicious circuit. Since only one trigger needs to be found to detect the HWT, we only need to have one entry in the LUT of the auxiliary circuit.

% When the LUT contains the correct trigger and payload, the functionality of TroLL is reproduced. However, SITP is applicable not only to TroLL but also to other types of HWT in general. This is because we only need to find 
\begin{figure}[htpb]
    \centering
    % \vspace{-0.1in}
    \includegraphics[width=.48\textwidth]{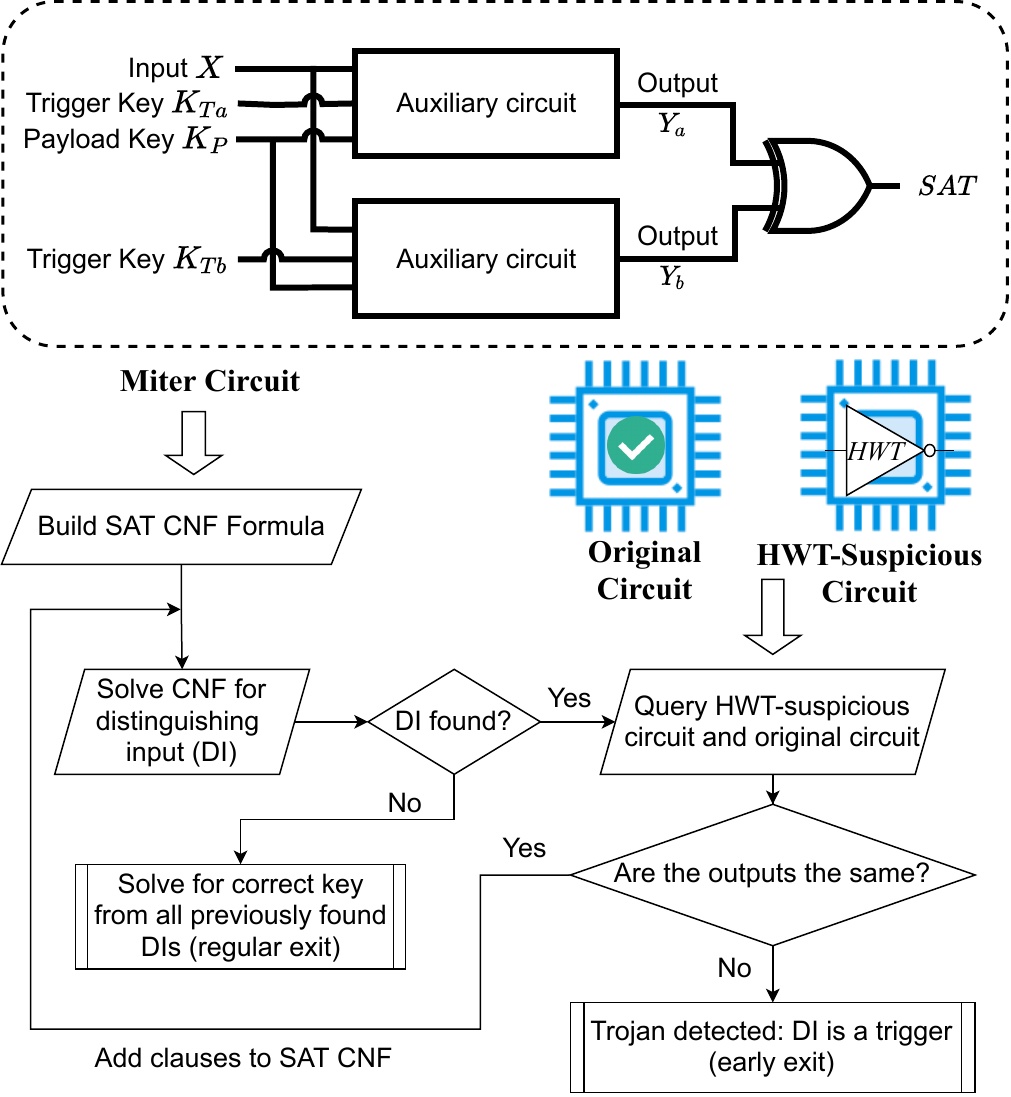}
    \vspace{-1mm}
    \caption{The SAT-based HWT Detection Flow}
    \vspace{-3mm}
    \label{fig:sat_detect}
\end{figure}

% \vspace{-2mm}
\subsubsection{Detection Flow}
% \vspace{-1mm}
The flow of SAT-based Detection is laid out in Fig. \ref{fig:sat_detect}. Similar to the SAT-based attack against logic locking introduced in Section \ref{ssec:locking}, a miter circuit is built using two copies of the auxiliary circuit and their outputs are XOR'ed. 
Let $F(\Vec{X})$ be Boolean function of the original circuit, $F_A(\Vec{X}, \Vec{K}_T, \Vec{K}_P)$ be that of the auxiliary circuit, and $H(\Vec{X})$ be that of the HWT-suspicious circuit. In the first iteration, the following SAT formula is solved to obtain the distinguishing input (DI):
\begin{equation}
     F_A(\Vec{DI_1}, \Vec{K}_{Ta}, \Vec{K}_P) \ne F_A(\Vec{DI_1}, \Vec{K}_{Tb}, \Vec{K}_P)
     \label{eq:iter1}
\end{equation}
The subscript of $DI$ stands for the iteration number. Then, both the original circuit and the HWT-suspicious circuit are queried with the DI. If the results are not equal, \ie $F(DI_1)\ne H(DI_1)$, then the HWT is detected and $DI_1$ is an HWT trigger. If they are equal, then let $O_1=H(DI_1)$. In the second iteration, clauses are added to ensure that the new keys found should produce correct output for $DI_1$ since it is not the trigger:
\begin{equation}
\begin{aligned}
     & F_A(\Vec{DI_2}, \Vec{K}_{Ta}, \Vec{K}_P) \ne F_A(\Vec{DI_2}, \Vec{K}_{Tb}, \Vec{K}_P) \\
     \bigwedge & F_A(\Vec{DI_1}, \Vec{K}_{Ta}, \Vec{K}_P) = F_A(\Vec{DI_1}, \Vec{K}_{Tb}, \Vec{K}_P) = O_1
     \label{eq:iter2}    
\end{aligned}
\end{equation}
The added clause will exclude \textit{any} trigger key $K_T$ that mistakes a non-trigger $\Vec{DI_1}$ as a trigger, which makes SAT-based detection potentially more efficient than purely testing-based detection approaches which only determine whether the test pattern is an HWT trigger or not.
% will only rule out one test pattern as the trigger at a time.
% In each iteration, a distinguishing input (DI) is solved for and new SAT clauses are added. These new clauses ensure that the trigger and payload keys found in the following iterations must
\begin{figure*}[bht]
    \centering
    \includegraphics[width=\textwidth, trim ={0.1cm 0.1cm 0.1cm 0.1cm}, clip]{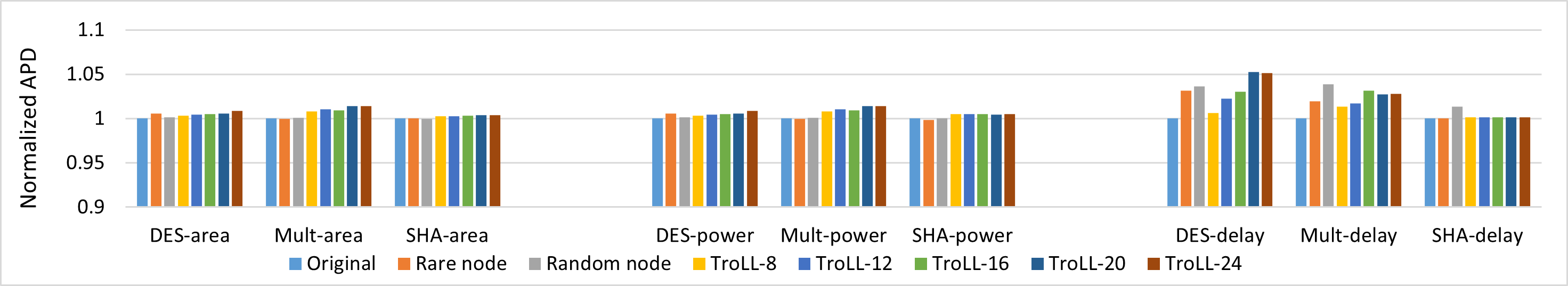}
    % \vspace{-0.2in}
    \caption{Mean area, power, and delay overhead of each HWT type}
    \label{fig:apd}
        % \vspace{-0.2in}
\end{figure*}

The process of SAT-based detection have some key differences from the SAT attack on logic locking: 
\begin{itemize}
    \item The oracle used in the formulation is the HWT-suspicious circuit under detection, instead of an activated chip.
    \item An early exit condition is added. If the DI produce a different output on the HWT-suspicious circuit compared to the original circuit, the detection process will terminate because an HWT is detected.
    \item The same payload key is applied to both copies of the auxiliary circuits to ensure that the output difference of the two copies is caused by the trigger key. 
    \item The correct key found by SAT attack on logic locking will make the locked circuit have the same functionality as the original circuit, whereas the SAT-based detection is not meant for replicating the exact functionality of the HWT-free circuit.
\end{itemize}

\subsection{Applicability of Netlist-based Attacks on Logic Locking for TroLL Detection}
There are attacks on logic locking that analyze the locked netlist for structural and functional information and achieve better results than the SAT attack. For example, Sirone et. al. proposed netlist analysis methodology to identify the MU's in SFLL and extracted the correct keys \cite{sirone2020functional}. The work of Han et. al. automated such analysis using Electronic Design Automation (EDA) tools and was applicable more generally to most logic locking methods based on the corrupt-and-correct principle \cite{han2021does}. These attacks are predicated on the attacker having access to the locked netlist of the circuit. From the netlist, the attacker can exploit structural and functional properties of the underlying logic locking approach to infer the correct key or to identify vulnerabilities.

In contrast, our threat model assumes that the defender (e.g., the design house) does not have access to the detailed netlist of the fabricated chip, especially if the fabrication process is outsourced to an untrusted foundry. 
Although our TroLL design borrows techniques from logic locking, the ``roles'' played the attacker (foundry) and defender (designer) are reversed. In logic locking, the designer inserts key gates and provides a locked netlist to an untrusted foundry. The foundry knows the modified netlist and tries to uncover the key. 
By contrast, in foundry-inserted hardware Trojans (HWT), the foundry secretly modifies the designer’s netlist, and the designer generally does not know the final netlist. 
If the designer wants to obtain the netlist of designs with Trojans inserted during fabrication, reverse-engineering the finished chip is required. This is a costly, destructive, and slow process \cite{torrance2011state}. On the other hand, if reverse engineering were performed, simple netlist comparison would reveal any deviation, making the attack unattractive to an adversary intent on stealth. Therefore, the mainstream HWT literature does not consider the defender to have access to the netlist \cite{saha2015improved, pan2021automated, pomeranz2004measure, chakraborty2009mero, lyu2020scalable}.
We summarize the threat models considered in logic locking and HWTs side-by-side in Table \ref{tab:threat_model} for a clearer comparison. 

\begin{table}[ht]
    \centering
    \caption{Contrasting threat models for logic locking and foundry-inserted hardware Trojans.}
    \begin{tabular}{@{}lcc@{}}
        \toprule
        & \textbf{Logic Locking} & \textbf{ HWTs} \\
        \midrule
        Netlist modified by            & Designer & Foundry \\
        Who tries to {decipher} the change? & Foundry & Designer \\
        Designer knows final netlist?  & Yes & No \\
        Foundry knows original netlist?& No  & Yes \\
        \bottomrule
    \end{tabular}
     \vspace{1mm}
    \label{tab:threat_model}
\end{table}
In summary, although TroLL leverages logic-locking {structures}, its attacker/defender assumptions align with the established HWT threat model, not with logic locking’s. 
This lack of access precludes the application of netlist-based attacks like \cite{sirone2020functional, han2021does} for Trojan detection. Instead, defenders typically rely on test-based approaches to detect anomalies in circuit behavior.

\subsection{Summary}
In this section, we introduce two types of novel HWT detection techniques that have potentials to detect TroLL more effectively than existing approaches. The evolved ATPG-based detection aims at finding the trigger based on TroLL's trigger selection algorithm, whereas SAT-based detection is an effort to take advantage of TroLL's resemblance to logic locking. We also clarify that netlist-based attacks on logic locking, albeit being more effective attacks, cannot be adapted to detect locking-based Trojans due to the lack of netlist access.
In the next section, we will examine the customized detection techniques alongside the existing ones to evaluate TroLL's ability to evade detection.
% We develop this SAT-based detection technique in an effort to take advantage of TroLL's resemblance to logic locking. It is noteworthy that the main intellectual contribution of this paper is TroLL and it is not our intention to promote this technique as a new HWT detection technique. In the next section, we will use SAT-based detection alongside ATPG-based and random sampling-based detection technique to evaluate TroLL's ability to evade detection and compare with existing types of HWT's.}

%unlike the SAT-based attacks on logic locking, this formulation is not meant for replicating the exact functionality of the HWT-infested circuit. Instead, it allows for the search of an HWT trigger pattern in an iterative manner and can terminate when any trigger is found. 

%an input pattern that cause any difference between the outputs of the original circuit and the HWT-suspicious circuit.

% \vspace{-3mm}
\section{Experiments} \label{sec:exp}
% \vspace{-1mm}
In this section, we present details on TroLL implementation and evaluation. We also compare the detection approaches introduced in Section \ref{sec:troll_detect} with existing state-of-the-art ATPG-based HWT detection approaches and random sampling on both TroLL and conventional HWT's.

\subsection{Experiment Setup}
In our experiments, we implement both TroLL and conventional hardware Trojans, including rare node triggered Trojans and random node triggered Trojans.  Three circuit benchmarks are used for the evaluation: DES, a 32-bit multiplier, and SHA-256, with a range of sizes as shown in Table \ref{tab:benchmarks}.
\begin{table}[tb]
    \centering
    \small
    % \vspace{-1mm}
    \caption{Benchmarks Used in HWT Evaluations}
        % \vspace{-2mm}
    \begin{tabular}{|c|c|c|c|}
    \hline
       Benchmark Name & \# Gates & \# Inputs & \# Outputs\\ \hline \hline
        %Reed-Solomon Encoder & 260\\ \hline
        %I2C & 1099 & 156 & 104\\ \hline
        DES & 6,473 & 256 & 245\\ \hline
        32-Bit Multiplier & 10,609 & 64 & 64\\ \hline
        SHA-256 & 51,222 & 512 & 256\\ \hline
        \multicolumn{2}{c}{}\\[-2ex]
    \end{tabular}
    \label{tab:benchmarks}
        % \vspace{-3mm}
\end{table}
Each benchmark is in gate-level Verilog format and all HWT's are incorporated by modifying the Verilog file. We use Icarus Verilog as the simulation tool. In Section \ref{ssec:trojan_implementation}, we describe which Trojans instances are made and their cost in terms of area, power and delay.

For HWT detection, we emulate post-fabrication testing by simulating the HWT-infested circuit and comparing the observable nodes, including output pins and flip-flop values, with the correct values. The simulation tool is also Icarus Verilog. This is consistent with real chip testing where the output pins can be read directly, the flip-flop values can be read through the scan chain, and other internal nodes inside combinational logic cones cannot be read directly. In Section \ref{ssec:detect_rare}, we present HWT detection results using multiple test-based methods.

\subsection{Trojan Implementation and Overhead}
\label{ssec:trojan_implementation}

We analyze and determine the rare values and associated probability of each internal node by simulating 100,000 randomly generated input patterns on each benchmark circuit. For rare node triggered Trojans, the triggers are selected directly based on this analysis. For TroLL, we choose trigger patterns using Algorithm 1 introduced in Section \ref{ssec:trigger_select}. Notice that the length of these triggers are the same as the circuit's input. When a shorter trigger length is needed, we choose a random subset of bits from the trigger patterns. 
For the HWT payload, we choose a subset of output pins to flip when the trigger condition is satisfied and the payload is the same across all the HWT instances for the same benchmark. 
Incorporating the MU output into the primary outputs guarantees that any corruption introduced is directly reflected in the observable behavior, minimizing the risk of the malicious effect being masked or rendered latent during internal signal propagation, which is desirable for the attacker. To improve the stealthiness of TroLL, the attacker can increase the trigger length, which will significantly decrease the likelihood of the trigger being present in the test patterns for Trojan detection. 

TroLL with 8, 12, 16, 20, and 24 bits trigger are implemented. For each type of HWT, we create 100 instances. The area, power and delay (APD) values are evaluated through synthesis in a 45nm library using Synopsys Design Compiler. The average APD values of each benchmark is shown in Figure \ref{fig:apd}. The APD values are normalized using those of the original design of each benchmark. 
For example, a value of 1.01 indicate a 1\% overhead.
As we can observe from the figures, the overhead percentage of TroLL is very low, especially for the largest benchmark (SHA-256). Hence, TroLL can be very well hidden in large designs and it is very difficult to identify them by APD analysis. This also underscores the scalability of TroLL.

\subsection{HWT Detection} \label{ssec:detect_rare}
\begin{figure}[t]
    \centering
    % \vspace{-1mm}
    \includegraphics[width=.48\textwidth, trim={0.1cm 0.5mm 0.1cm 0.5mm}, clip]{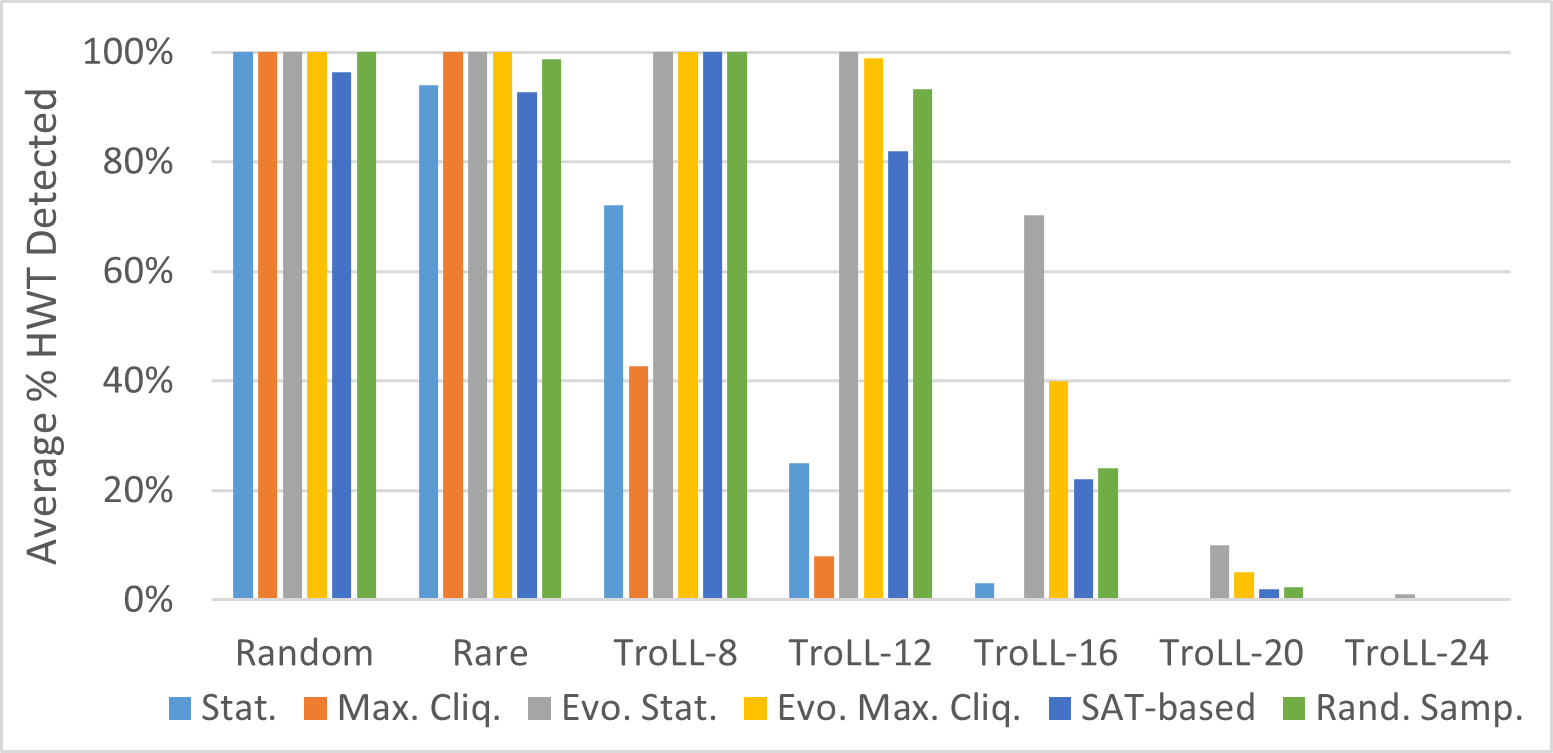}
    % \vspace{-0.2in}
    \caption{Comparison of HWT Detection Approaches}
    \vspace{-2mm}
    \label{fig:comp}
\end{figure}

% \vspace{-2mm}
In this section, we evaluate the effectiveness of HWT detection techniques: ATPG-based methods, including statistical test generation (MERO~\cite{chakraborty2009mero}) and maximal clique sampling (TARMAC~\cite{lyu2020scalable}), the evolved versions of statistical test generation and maximal clique sampling, SAT-based detection, and random sampling.
Figure \ref{fig:comp} provides an overall comparison of the efficacy of these detection approaches. In the figure, the percentage of detected HWTs is clustered by the HWT type, including rare node Trojans, random node Trojans, and TroLL with 8 to 24 bit triggers.
For a certain HWT type and size, the percentage of HWTs detected is calculated by dividing the number of HWTs detected by the total number of HWTs of the same type and size.
In general, the detected percentage of any approach drops significantly as TroLL's trigger length increases. This underscores TroLL's resilience to test-based HWT detection approaches.
We can also see the performance difference among the detection approaches. Overall, the conventional ATPG-based HWT detection approaches have the worst efficacy against TroLL. However, their evolved versions, especially evolved statistical sampling, outperform all other methods on TroLL with 12 to 20 bits in the trigger. The detection percentage of random sampling and SAT-based detection lie between the original and evolved ATPG-based detection approaches.

\begin{table*}[t]
\scriptsize
\caption{Hardware Trojan detection performance by existing and proposed methods}
% % \vspace{-0.1in}
\centering
\begin{tabular}{|c|c|c|c|c|c|c|c|c|c|c|c|c|c|c|}
\hline
\multirow{3}{*}{\textbf{Benchmark}} & \multicolumn{7}{c|}{\textbf{Statistical Test Generation~\cite{chakraborty2009mero}}} & \multicolumn{7}{c|}{\textbf{Maximal Clique Sampling~\cite{lyu2020scalable}}} \\ \cline{2-15} 
& \multirow{2}{*}{\textbf{Random}} & \multirow{2}{*}{\textbf{Rare}} & \multicolumn{5}{c|}{\textbf{TroLL with trigger length}} & \multirow{2}{*}{\textbf{Random}} & \multirow{2}{*}{\textbf{Rare}} & \multicolumn{5}{c|}{\textbf{TroLL with trigger length}}\\ \cline{4-8}\cline{11-15}
&&& \textbf{8} & \textbf{12} & \textbf{16} & \textbf{20} & \textbf{24} &&& \textbf{8} & \textbf{12} & \textbf{16} & \textbf{20} & \textbf{24} \\ \hline
\textbf{DES} & \multicolumn{1}{c|}{100\%} & \multicolumn{1}{c|}{100\%} & \multicolumn{1}{c|}{38\%} & \multicolumn{1}{c|}{5\%} & 0\% & 0\% & 0\%  &  \multicolumn{1}{c|}{100\%} & \multicolumn{1}{c|}{100\%} & \multicolumn{1}{c|}{46\%} & \multicolumn{1}{c|}{8\%} & 0\% & 0\% & 0\%  \\ \hline
\textbf{Mult-32} & \multicolumn{1}{c|}{100\%} & \multicolumn{1}{c|}{83\%} & \multicolumn{1}{c|}{100\%} & \multicolumn{1}{c|}{62\%} &  \multicolumn{1}{c|}{8\%}& 1\% & 0\% & \multicolumn{1}{c|}{100\%} & \multicolumn{1}{c|}{100\%} & \multicolumn{1}{c|}{55\%} & \multicolumn{1}{c|}{15\%} & 1\% & 0\% & 0\%\\ \hline
\textbf{SHA-256} & \multicolumn{1}{c|}{100\%} & \multicolumn{1}{c|}{100\%} & \multicolumn{1}{c|}{77\%} & \multicolumn{1}{c|}{9\%} & 1\% & 0\% & 0\% &  \multicolumn{1}{c|}{100\%} & \multicolumn{1}{c|}{100\%} & \multicolumn{1}{c|}{27\%} & \multicolumn{1}{c|}{0\%} & 0\% & 0\% & 0\%\\ \hline
%%%%%%%%%%%%%%%%%%%%%%%%%%%%%%%%%%%%%%%%%%%%%%%%%%%%%%%%%%%%%%%%%%%%%%%%%%%%%%%%%%%%%%%%%%%%%%%%%%%%%%%%%%%%%%%%%%%%%%%%
\hline
\multirow{3}{*}{\textbf{Benchmark}} & \multicolumn{7}{c|}{\textbf{Evolved Statistical Test Generation}} & \multicolumn{7}{c|}{\textbf{Evolved Maximal Clique Sampling}} \\ \cline{2-15} 
& \multirow{2}{*}{\textbf{Random}} & \multirow{2}{*}{\textbf{Rare}} & \multicolumn{5}{c|}{\textbf{TroLL with trigger length}} & \multirow{2}{*}{\textbf{Random}} & \multirow{2}{*}{\textbf{Rare}} & \multicolumn{5}{c|}{\textbf{TroLL with trigger length}}\\ \cline{4-8}\cline{11-15}
&&& \textbf{8} & \textbf{12} & \textbf{16} & \textbf{20} & \textbf{24} &&& \textbf{8} & \textbf{12} & \textbf{16} & \textbf{20} & \textbf{24} \\ \hline
\textbf{DES} & \multicolumn{1}{c|}{100\%} & \multicolumn{1}{c|}{100\%} & \multicolumn{1}{c|}{100\%} & \multicolumn{1}{c|}{100\%} & 59\% & 11\% & 1\%  &  \multicolumn{1}{c|}{100\%} & \multicolumn{1}{c|}{100\%} & \multicolumn{1}{c|}{100\%} & \multicolumn{1}{c|}{99\%} & 45\% & 6\% & 0\%  \\ \hline
\textbf{Mult-32} & \multicolumn{1}{c|}{100\%} & \multicolumn{1}{c|}{99\%} & \multicolumn{1}{c|}{100\%} & \multicolumn{1}{c|}{100\%} &  \multicolumn{1}{c|}{85\%}& 10\% & 0\%  & \multicolumn{1}{c|}{100\%} & \multicolumn{1}{c|}{100\%} & \multicolumn{1}{c|}{100\%} & \multicolumn{1}{c|}{99\%} & 41\% & 4\% & 0\%\\ \hline
\textbf{SHA-256} & \multicolumn{1}{c|}{100\%} & \multicolumn{1}{c|}{100\%} & \multicolumn{1}{c|}{100\%} & \multicolumn{1}{c|}{100\%} & 67\% & 8\% & 1\% & \multicolumn{1}{c|}{100\%} & \multicolumn{1}{c|}{100\%} & \multicolumn{1}{c|}{100\%} & \multicolumn{1}{c|}{100\%} & 34\% & 4\% & 0\%\\ \hline
%%%%%%%%%%%%%%%%%%%%%%%%%%%%%%%%%%%%%%%%%%%%%%%%%%%%%%%%%%%%%%%%%%%%%%%%%%%%%%%%%%%%%%%%%%%%%%%%%%%%%%%%%%%%%%%%%%%%%%%%
\hline
\multirow{3}{*}{\textbf{Benchmark}} & \multicolumn{7}{c|}{\textbf{SAT-based Detection}} & \multicolumn{7}{c|}{\textbf{Random Sampling}}  \\ \cline{2-15} 
& \multirow{2}{*}{\textbf{Random}} & \multirow{2}{*}{\textbf{Rare}} & \multicolumn{5}{c|}{\textbf{TroLL with trigger length}} & \multirow{2}{*}{\textbf{Random}} & \multirow{2}{*}{\textbf{Rare}} & \multicolumn{5}{c|}{\textbf{TroLL with trigger length}} \\ \cline{4-8}\cline{11-15}
&&& \textbf{8} & \textbf{12} & \textbf{16} & \textbf{20} & \textbf{24} &&& \textbf{8} & \textbf{12} & \textbf{16} & \textbf{20} & \textbf{24}\\ \hline
\textbf{DES} & \multicolumn{1}{c|}{94\%} & \multicolumn{1}{c|}{84\%} & \multicolumn{1}{c|}{100\%} & \multicolumn{1}{c|}{100\%} & 26\% & 1\% & 0\% & \multicolumn{1}{c|}{100\%} & \multicolumn{1}{c|}{100\%} & \multicolumn{1}{c|}{100\%} & \multicolumn{1}{c|}{99\%} & 31\% & 1\% & 0\%\\ \hline
\textbf{Mult-32} & \multicolumn{1}{c|}{100\%} & \multicolumn{1}{c|}{96\%} & \multicolumn{1}{c|}{100\%} & \multicolumn{1}{c|}{82\%} &  \multicolumn{1}{c|}{23\%}& 4\% & 0\% & \multicolumn{1}{c|}{100\%} & \multicolumn{1}{c|}{96\%} & \multicolumn{1}{c|}{100\%} & \multicolumn{1}{c|}{95\%} & 10\% & 1\% & 0\%  \\ \hline 
\textbf{SHA-256} & \multicolumn{1}{c|}{95\%} & \multicolumn{1}{c|}{98\%} & \multicolumn{1}{c|}{100\%} & \multicolumn{1}{c|}{70\%} & 16\% & 1\% & 1\% & \multicolumn{1}{c|}{100\%} & \multicolumn{1}{c|}{100\%} & \multicolumn{1}{c|}{100\%} & \multicolumn{1}{c|}{86\%} & 33\% & 5\% & 0\% \\ \hline
%%%%%%%%%%%%%%%%%%%%%%%%%%%%%%%%%%%%%%%%%%%%%%%%%%%%%%%

\end{tabular}
\label{tab:conventional}
\label{tab:sitp}
\vspace{-2mm}
\end{table*}

Table \ref{tab:conventional} shows the percentage of HWT's detected using six test-based methods. 
The results of the original ATPG-based methods, including statistical methods \cite{chakraborty2009mero} and maximal clique sampling \cite{lyu2020scalable} are shown on the top section of Table \ref{tab:conventional}. For rare and random node triggered Trojans, these methods can detect 100\% of the HWT's in most cases. For TroLL, these detection methods are not effective as the detection percentage fades quickly with the increase in trigger length.
This is because TroLL's trigger selection algorithm, as presented in Section \ref{ssec:trigger_select}, intentionally avoids sensitizing any rare nodes within the original circuit. As both statistical test generation \cite{chakraborty2009mero} and maximal clique sampling \cite{lyu2020scalable} ensures that each test pattern will sensitize some rare nodes in the circuit, they are unlikely to sensitize the triggers of TroLL.

In the middle section of Table \ref{tab:conventional}, we show the HWT detection results with the evolved ATPG-based approaches. Compared to the original ATPG-based approaches, the evolved ones are able to detect more TroLL-type HWTs. 
Specifically, the improvement in detection percentage is most significant for TroLL with trigger length up to 20 bits. In these detection methods, thanks to the knowledge to TroLL's trigger selection mechanism, the defender can set the targeted node values to meet TroLL's trigger criteria when using ATPG algorithms to generate test patterns. This is the main cause of the improvement.

SAT-based detection is implemented based on the code framework of SAT-based attacks on logic locking presented in \cite{subramanyan2015evaluating}. We limit the time of each SAT-based detection run to 48 hours and a Trojan is considered as not detected if no trigger pattern is found within this time frame. In the bottom left division of Table \ref{tab:sitp}, we show the percentage of HWT detected by SAT for each benchmark and type of HWT.
The efficacy of this approach is better than statistical methods \cite{chakraborty2009mero} and maximal clique sampling \cite{lyu2020scalable} but not as good as their evolved counterpart. A possible reason is that the TroLL avoids using rare nodes as triggers. As these rare nodes values are inherently hard to satisfy, not using them improves the efficacy of SAT-based detection.

The random sampling detection results are shown in the bottom right division of Table \ref{tab:sitp}. This method can be used as the baseline case as it does not require any specialized algorithm. In Figure \ref{fig:comp}, it can be seen that the ATPG-based approaches evolved have higher efficacy than random sampling in TroLL, while their original versions perform worse than random sampling. This indicates that the customization of the ATPG-based approaches presented in Section \ref{ssec:custom_atpg} is effective against TroLL.

\subsection{Discussions} \label{sec:discussion}
Through the results presented in Section \ref{ssec:detect_rare}, we can see that, for each existing and proposed ATPG-based detection technique studied, there is a range of trigger bit numbers where the detection percentage drops from 100\% to 0. This trend reflects a fundamental limitation of test-based HWT detection approaches. Test-based approaches can detect an HWT only when the test input triggers the HWT and leads to an incorrect output. The input space is exponential in the number of input bits, while only a limited number of test patterns can be applied during the test time frame. Therefore, the test patterns account for a very small subset of the entire input space, forcing the ATPG algorithms to choose test patterns that will most likely trigger the HWT's. Such a likelihood can only be assumed from the HWT insertion algorithms, and any incorrect assumption will diminish the likelihood of successful HWT detection. This can explain the poor performance of using test patterns generated for rare node-triggered HWTs for TroLL. When the correct assumptions are used, as is the case for the evolved ATPG-based approaches for TroLL, the detection percentage becomes better. However, the detection is still test-based, and the test fraction of the input space still becomes exponentially smaller as the number of input bits increases. Therefore, it is unlikely that any purely test-based HWT detection approach can maintain its efficacy with the increase in input bits.

\section{Conclusion} \label{sec:conc}

In this paper, we present a novel type of Hardware Trojans based on logic locking, TroLL. TroLL is constructed by retaining the mutation unit (MU) and removing the restore unit (RU) of state-of-the-art logic locking techniques. The trigger patterns of TroLL are selected in a way that avoids sensitizing the internal rare signals of the original circuit, thereby evading state-of-the-art ATPG-based detection schemes.
In an attempt to formulate an effective detection approach against TroLL, we tried several different approaches, including evolving the ATPG-based approaches targeting the internal nodes' prevalent values in addition to the rare values, and adapting the SAT-based attacks on logic locking to HWT detection. We also use random sampling as a reference.
We found that the evolved ATPG-based approaches performed better than random sampling, but even these approaches' efficacy diminishes as TroLL's triggers get longer. 
Therefore, we have identified TroLL as a new threat to the integrity of hardware manufactured in untrusted fabrication facilities, and it is necessary to find a scalable detection approach against TroLL.

On a broader scale, this paper reminds us that even a design protection scheme (such as logic locking) can be a double edged sword. Meanwhile, just like the SAT attack can be turned to an HWT detection scheme, we can examine other attacks against logic locking in the search for a more effective detection approach against TroLL.

% \begin{acks}
%\section*{Acknowledgment}

\bibliographystyle{IEEEtran}
\bibliography{IEEEabrv,x_references}

\begin{thebibliography}{10}
\providecommand{\url}[1]{#1}
\csname url@samestyle\endcsname
\providecommand{\newblock}{\relax}
\providecommand{\bibinfo}[2]{#2}
\providecommand{\BIBentrySTDinterwordspacing}{\spaceskip=0pt\relax}
\providecommand{\BIBentryALTinterwordstretchfactor}{4}
\providecommand{\BIBentryALTinterwordspacing}{\spaceskip=\fontdimen2\font plus
\BIBentryALTinterwordstretchfactor\fontdimen3\font minus \fontdimen4\font\relax}
\providecommand{\BIBforeignlanguage}[2]{{%
\expandafter\ifx\csname l@#1\endcsname\relax
\typeout{** WARNING: IEEEtran.bst: No hyphenation pattern has been}%
\typeout{** loaded for the language `#1'. Using the pattern for}%
\typeout{** the default language instead.}%
\else
\language=\csname l@#1\endcsname
\fi
#2}}
\providecommand{\BIBdecl}{\relax}
\BIBdecl

\bibitem{chakraborty2019keynote}
A.~Chakraborty \emph{et~al.}, ``Keynote: A disquisition on logic locking,'' \emph{IEEE Transactions on CAD}, vol.~39, no.~10, pp. 1952--1972, 2019.

\bibitem{frey2015exploiting}
J.~Frey and Q.~Yu, ``Exploiting state obfuscation to detect hardware trojans in noc network interfaces,'' in \emph{MWSCAS}, 2015, pp. 1--4.

\bibitem{hu2019leveraging}
W.~Hu \emph{et~al.}, ``Leveraging unspecified functionality in obfuscated hardware for trojan and fault attacks,'' in \emph{AsianHOST}, 2019, pp. 1--6.

\bibitem{yu2017exploiting}
Q.~Yu \emph{et~al.}, ``Exploiting hardware obfuscation methods to prevent and detect hardware trojans,'' in \emph{MWSCAS}, 2017, pp. 819--822.

\bibitem{dupuis2014novel}
S.~Dupuis \emph{et~al.}, ``A novel hardware logic encryption technique for thwarting illegal overproduction and hardware trojans,'' in \emph{IOLTS}, 2014, pp. 49--54.

\bibitem{mirmohammadi2023new}
Z.~Mirmohammadi and S.~Etemadi~Borujeni, ``A new optimal method for the secure design of combinational circuits against hardware trojans using interference logic locking,'' \emph{Electronics}, vol.~12, no.~5, p. 1107, 2023.

\bibitem{cruz2022analysis}
J.~Cruz, P.~Gaikwad, and S.~Bhunia, ``Analysis of hardware trojan resilience enabled through logic locking,'' in \emph{2022 Asian hardware oriented security and trust symposium (AsianHOST)}.\hskip 1em plus 0.5em minus 0.4em\relax IEEE, 2022, pp. 1--6.

\bibitem{wang2024trolloc}
F.~Wang, Q.~Wang, L.~Alrahis, B.~Fu, S.~Jiang, X.~Zhang, O.~Sinanoglu, T.-Y. Ho, E.~F. Young, and J.~Knechtel, ``Trolloc: Logic locking and layout hardening for ic security closure against hardware trojans,'' \emph{arXiv preprint arXiv:2405.05590}, 2024.

\bibitem{chakraborty2009security}
R.~Chakraborty and S.~Bhunia, ``Security against hardware trojan through a novel application of design obfuscation,'' in \emph{ICCAD}, 2009, pp. 113--116.

\bibitem{maynard2024reconfigurable}
J.~Maynard and A.~Rezaei, ``Reconfigurable run-time hardware trojan mitigation for logic-locked circuits,'' in \emph{2024 IEEE 17th Dallas Circuits and Systems Conference (DCAS)}.\hskip 1em plus 0.5em minus 0.4em\relax IEEE, 2024, pp. 1--6.

\bibitem{chakraborty2009mero}
R.~S. Chakraborty, F.~Wolff, S.~Paul, C.~Papachristou, and S.~Bhunia, ``Mero: A statistical approach for hardware trojan detection,'' in \emph{International Workshop on Cryptographic Hardware and Embedded Systems}, 2009, pp. 396--410.

\bibitem{lyu2020scalable}
Y.~Lyu and P.~Mishra, ``Scalable activation of rare triggers in hardware trojans by repeated maximal clique sampling,'' \emph{IEEE Transactions on CAD}, 2020.

\bibitem{zhang2013hardware}
J.~Zhang and Q.~Xu, ``On hardware trojan design and implementation at register-transfer level,'' in \emph{2013 IEEE international symposium on hardware-oriented security and trust (HOST)}.\hskip 1em plus 0.5em minus 0.4em\relax IEEE, 2013, pp. 107--112.

\bibitem{tsoutsos2014advanced}
N.~G. Tsoutsos, C.~Konstantinou, and M.~Maniatakos, ``Advanced techniques for designing stealthy hardware trojans,'' in \emph{2014 51st ACM/EDAC/IEEE Design Automation Conference (DAC)}.\hskip 1em plus 0.5em minus 0.4em\relax IEEE, 2014, pp. 1--4.

\bibitem{fern2016hiding}
N.~Fern, I.~San, C.~K. Ko{\c{c}}, and K.-T.~T. Cheng, ``Hiding hardware trojan communication channels in partially specified soc bus functionality,'' \emph{IEEE Transactions on Computer-Aided Design of Integrated Circuits and Systems}, vol.~36, no.~9, pp. 1435--1444, 2016.

\bibitem{jin2009experiences}
Y.~Jin, N.~Kupp, and Y.~Makris, ``Experiences in hardware trojan design and implementation,'' in \emph{2009 IEEE International Workshop on Hardware-Oriented Security and Trust}.\hskip 1em plus 0.5em minus 0.4em\relax IEEE, 2009, pp. 50--57.

\bibitem{reece2012stealth}
T.~Reece, D.~B. Limbrick, X.~Wang, B.~T. Kiddie, and W.~H. Robinson, ``Stealth assessment of hardware trojans in a microcontroller,'' in \emph{2012 IEEE 30th International Conference on Computer Design (ICCD)}.\hskip 1em plus 0.5em minus 0.4em\relax IEEE, 2012, pp. 139--142.

\bibitem{sturton2011defeating}
C.~Sturton, M.~Hicks, D.~Wagner, and S.~T. King, ``Defeating uci: Building stealthy and malicious hardware,'' in \emph{2011 IEEE symposium on security and privacy}.\hskip 1em plus 0.5em minus 0.4em\relax IEEE, 2011, pp. 64--77.

\bibitem{xue2020ten}
M.~Xue, C.~Gu, W.~Liu, S.~Yu, and M.~O'Neill, ``Ten years of hardware trojans: a survey from the attacker's perspective,'' \emph{IET Computers \& Digital Techniques}, vol.~14, no.~6, pp. 231--246, 2020.

\bibitem{karri2010trustworthy}
R.~Karri, J.~Rajendran, K.~Rosenfeld, and M.~Tehranipoor, ``Trustworthy hardware: Identifying and classifying hardware trojans,'' \emph{Computer}, vol.~43, no.~10, pp. 39--46, 2010.

\bibitem{rathmair2014applied}
M.~Rathmair, F.~Schupfer, and C.~Krieg, ``Applied formal methods for hardware trojan detection,'' in \emph{2014 IEEE International Symposium on Circuits and Systems (ISCAS)}.\hskip 1em plus 0.5em minus 0.4em\relax IEEE, 2014, pp. 169--172.

\bibitem{fern2017detecting}
N.~Fern, I.~San, and K.-T.~T. Cheng, ``Detecting hardware trojans in unspecified functionality through solving satisfiability problems,'' in \emph{2017 22nd Asia and South Pacific Design Automation Conference (ASP-DAC)}.\hskip 1em plus 0.5em minus 0.4em\relax IEEE, 2017, pp. 598--504.

\bibitem{bao2017reverse}
C.~Bao, Y.~Xie, Y.~Liu, and A.~Srivastava, ``Reverse engineering-based hardware trojan detection,'' \emph{The Hardware Trojan War: Attacks, Myths, and Defenses}, p. 269, 2017.

\bibitem{vashistha2021detecting}
N.~Vashistha, H.~Lu, Q.~Shi, D.~L. Woodard, N.~Asadizanjani, and M.~Tehranipoor, ``Detecting hardware trojans using combined self testing and imaging,'' \emph{IEEE Transactions on Computer-Aided Design of Integrated Circuits and Systems}, 2021.

\bibitem{torrance2011state}
R.~Torrance and D.~James, ``The state-of-the-art in semiconductor reverse engineering,'' in \emph{Proceedings of the 48th Design Automation Conference}, 2011, pp. 333--338.

\bibitem{waite2021preparation}
A.~R. Waite, Y.~Patel, J.~J. Kelley, J.~H. Scholl, J.~Baur, A.~Kimura, E.~D. Udelhoven, G.~D. Via, R.~Ott, and D.~L. Brooks, ``Preparation, imaging, and design extraction of the front-end-of-line and middle-of-line in a 14 nm node finfet device,'' in \emph{2021 IEEE Physical Assurance and Inspection of Electronics (PAINE)}.\hskip 1em plus 0.5em minus 0.4em\relax IEEE, 2021, pp. 1--6.

\bibitem{krachenfels2023trojan}
T.~Krachenfels, J.-P. Seifert, and S.~Tajik, ``Trojan awakener: detecting dormant malicious hardware using laser logic state imaging (extended version),'' \emph{Journal of Cryptographic Engineering}, vol.~13, no.~4, pp. 485--499, 2023.

\bibitem{TSMC_Q2_2025_transcript}
{Taiwan Semiconductor Manufacturing Company Limited}, ``Second-quarter 2025 earnings conference call transcript,'' July 2025, n3 reported as 24\% of wafer revenue; accessed 5 August 2025.

\bibitem{Samsung_3nm_press_2022}
{Samsung Electronics Co.\, Ltd.}, ``Samsung begins chip production using 3 nm process technology with gaa architecture,'' Press release, June 2022, accessed 5 August 2025.

\bibitem{cruz2018automated}
J.~Cruz, Y.~Huang, P.~Mishra, and S.~Bhunia, ``An automated configurable trojan insertion framework for dynamic trust benchmarks,'' in \emph{2018 Design, Automation \& Test in Europe Conference \& Exhibition (DATE)}.\hskip 1em plus 0.5em minus 0.4em\relax IEEE, 2018, pp. 1598--1603.

\bibitem{abbassi2019trojanzero}
I.~H. Abbassi, F.~Khalid, S.~Rehman, A.~M. Kamboh, A.~Jantsch, S.~Garg, and M.~Shafique, ``Trojanzero: Switching activity-aware design of undetectable hardware trojans with zero power and area footprint,'' in \emph{2019 Design, Automation \& Test in Europe Conference \& Exhibition (DATE)}.\hskip 1em plus 0.5em minus 0.4em\relax IEEE, 2019, pp. 914--919.

\bibitem{pomeranz2004measure}
I.~Pomeranz and S.~M. Reddy, ``A measure of quality for n-detection test sets,'' \emph{IEEE Trans. on Computers}, vol.~53, no.~11, pp. 1497--1503, 2004.

\bibitem{saha2015improved}
S.~Saha, R.~S. Chakraborty, S.~S. Nuthakki, D.~Mukhopadhyay \emph{et~al.}, ``Improved test pattern generation for hardware trojan detection using genetic algorithm and boolean satisfiability,'' in \emph{International Workshop on Cryptographic Hardware and Embedded Systems}.\hskip 1em plus 0.5em minus 0.4em\relax Springer, 2015, pp. 577--596.

\bibitem{pan2021automated}
Z.~Pan and P.~Mishra, ``Automated test generation for hardware trojan detection using reinforcement learning,'' in \emph{Asia and South Pacific Design Automation Conference}, 2021, pp. 408--413.

\bibitem{subramanyan2015evaluating}
P.~Subramanyan, S.~Ray, and S.~Malik, ``Evaluating the security of logic encryption algorithms,'' in \emph{Hardware Oriented Security and Trust (HOST), 2015 IEEE International Symposium on}.\hskip 1em plus 0.5em minus 0.4em\relax IEEE, 2015, pp. 137--143.

\bibitem{xie2016mitigating}
Y.~Xie and A.~Srivastava, ``Mitigating sat attack on logic locking,'' in \emph{International Conference on Cryptographic Hardware and Embedded Systems}.\hskip 1em plus 0.5em minus 0.4em\relax Springer, 2016, pp. 127--146.

\bibitem{yasin2016sarlock}
M.~Yasin, B.~Mazumdar, J.~J. Rajendran, and O.~Sinanoglu, ``Sarlock: Sat attack resistant logic locking,'' in \emph{Hardware Oriented Security and Trust (HOST), 2016 IEEE International Symposium on}.\hskip 1em plus 0.5em minus 0.4em\relax IEEE, 2016, pp. 236--241.

\bibitem{yasin2017provably}
M.~Yasin, A.~Sengupta, M.~T. Nabeel, M.~Ashraf, J.~J. Rajendran, and O.~Sinanoglu, ``Provably-secure logic locking: From theory to practice,'' in \emph{Proceedings of the 2017 ACM SIGSAC Conference on Computer and Communications Security}.\hskip 1em plus 0.5em minus 0.4em\relax ACM, 2017, pp. 1601--1618.

\bibitem{liu2021robust}
Y.~Liu, M.~Zuzak, Y.~Xie, A.~Chakraborty, and A.~Srivastava, ``Robust and attack resilient logic locking with a high application-level impact,'' \emph{ACM Journal on Emerging Technologies in Computing Systems (JETC)}, vol.~17, no.~3, pp. 1--22, 2021.

\bibitem{shakya2020cas}
B.~Shakya, X.~Xu, M.~Tehranipoor, and D.~Forte, ``Cas-lock: A security-corruptibility trade-off resilient logic locking scheme,'' \emph{IACR Transactions on Cryptographic Hardware and Embedded Systems}, pp. 175--202, 2020.

\bibitem{liu2020strong}
Y.~Liu, M.~Zuzak, Y.~Xie, A.~Chakraborty, and A.~Srivastava, ``Strong anti-sat: Secure and effective logic locking,'' in \emph{2020 21st International Symposium on Quality Electronic Design (ISQED)}.\hskip 1em plus 0.5em minus 0.4em\relax IEEE, 2020, pp. 199--205.

\bibitem{creignou1996complexity}
N.~Creignou and M.~Hermann, ``Complexity of generalized satisfiability counting problems,'' \emph{Information and computation}, vol. 125, no.~1, pp. 1--12, 1996.

\bibitem{yasin2017security}
M.~Yasin, B.~Mazumdar, O.~Sinanoglu, and J.~Rajendran, ``Security analysis of anti-sat,'' in \emph{2017 22nd Asia and South Pacific Design Automation Conference (ASP-DAC)}.\hskip 1em plus 0.5em minus 0.4em\relax IEEE, 2017, pp. 342--347.

\bibitem{mohyuddin2011probabilistic}
N.~Mohyuddin, E.~Pakbaznia, and M.~Pedram, ``Probabilistic error propagation in a logic circuit using the boolean difference calculus,'' in \emph{Advanced Techniques in Logic Synthesis, Optimizations and Applications}.\hskip 1em plus 0.5em minus 0.4em\relax Springer, 2011, pp. 359--381.

\bibitem{sirone2020functional}
D.~Sirone and P.~Subramanyan, ``Functional analysis attacks on logic locking,'' \emph{IEEE Transactions on Information Forensics and Security}, vol.~15, pp. 2514--2527, 2020.

\bibitem{han2021does}
Z.~Han, M.~Yasin, and J.~J. Rajendran, ``Does logic locking work with $\{$EDA$\}$ tools?'' in \emph{30th USENIX Security Symposium (USENIX Security 21)}, 2021, pp. 1055--1072.

\end{thebibliography}

% \appendix
\begin{IEEEbiography}[{\includegraphics[width=1in]{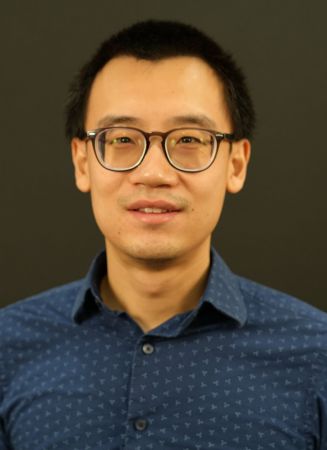}}]{Yuntao Liu} is an Assistant Professor in the Department of Electrical and Computer Engineering at Lehigh University. His research interests include hardware supply chain security, hardware security for artificial intelligence, side-channel analysis, quantum computing security, and biocomputing modeling and security. He received his PhD in Electrical Engineering from the University of Maryland, College Park in 2020.
% and B.Eng. from the College of Electrical Engineering, Zhejiang University, China in 2014.
\end{IEEEbiography}

\begin{IEEEbiography}[{\includegraphics[width=1in]{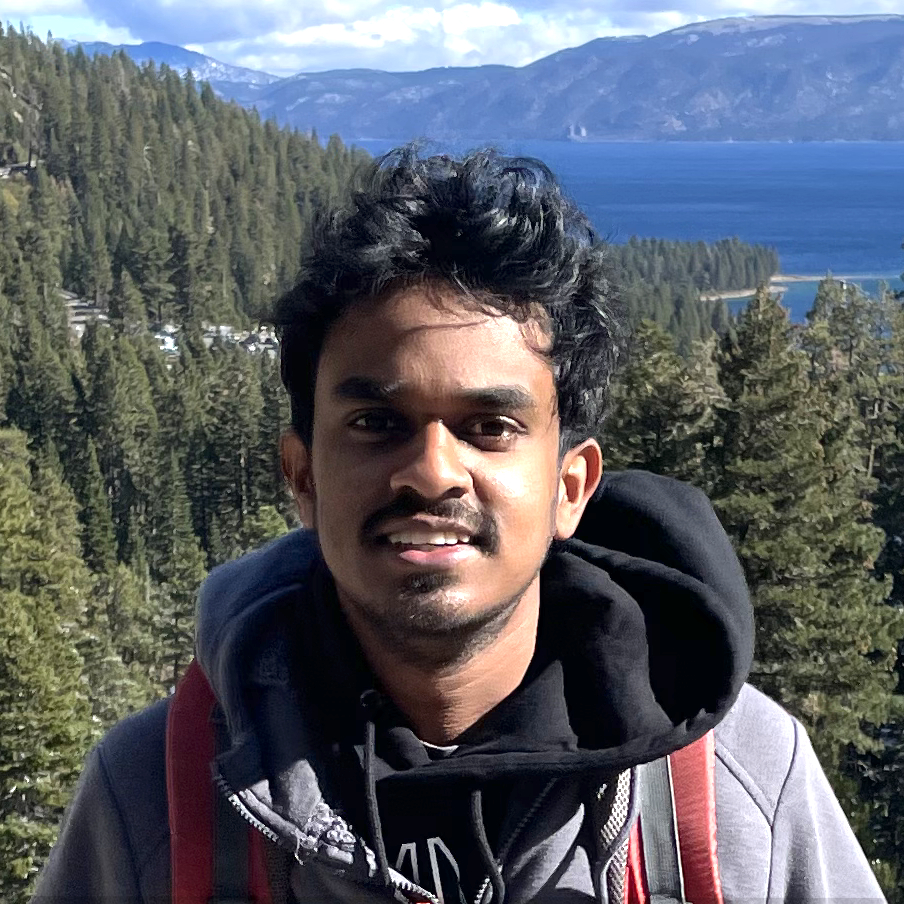}}]{Aruna Jayasena} is an Assistant Professor in the Department of Computer Science and Engineering at the University of Tennessee at Chattanooga. His research interests include systems security, hardware-firmware co-validation, applied cryptography, trusted execution, side-channel analysis, and test generation. He received his PhD in Computer Engineering from the University of Florida in 2025.
% and B.S. in the Department of Computer Science and Engineering at the University of Moratuwa, Sri Lanka in 2019. 
\end{IEEEbiography}

\begin{IEEEbiography}[{\includegraphics[width=1in]{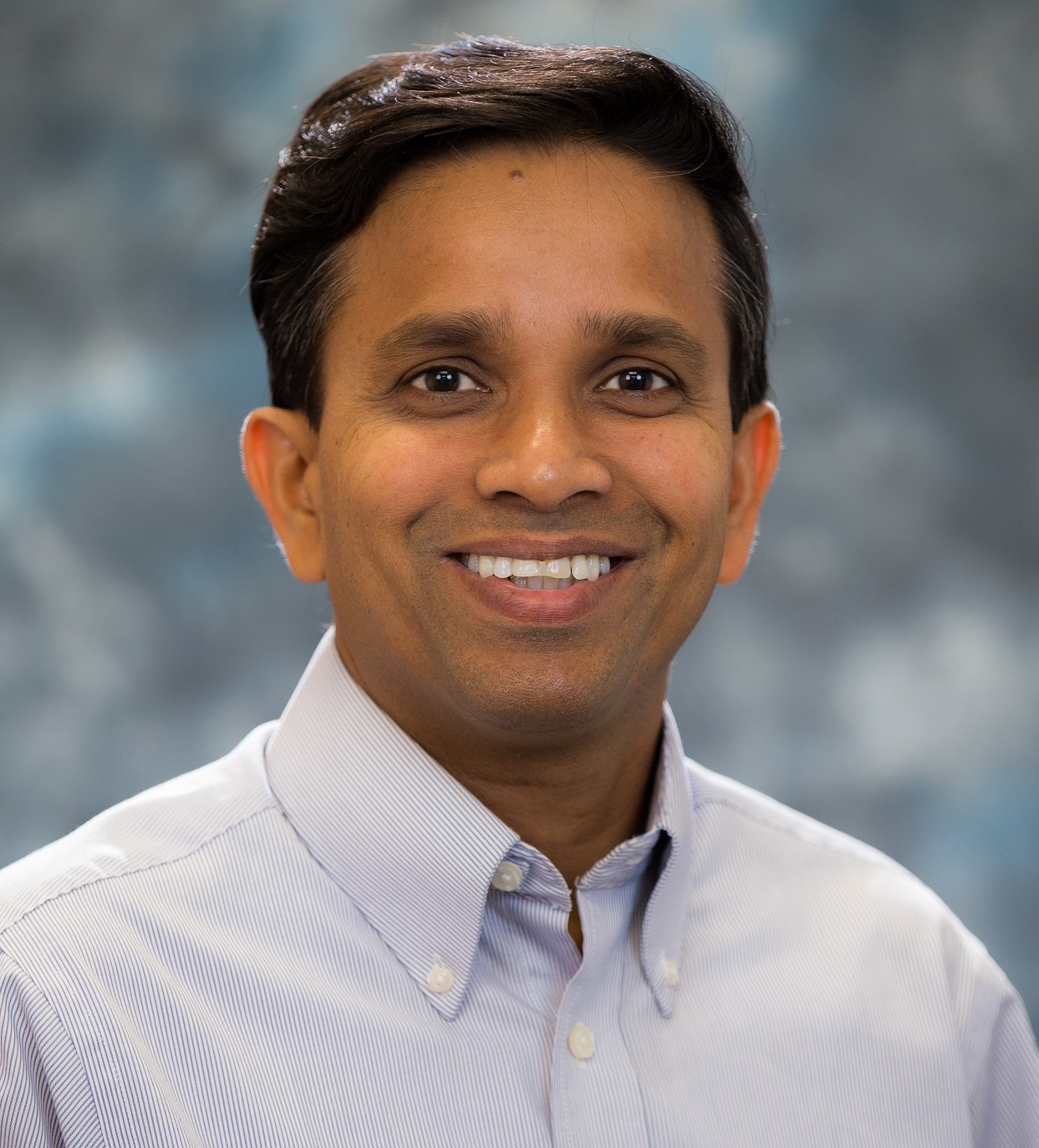}}]{Prabhat Mishra} is a Professor in the Department of Computer and Information Science and Engineering at the University of Florida. His research interests include embedded systems, hardware security, trustworthy AI, and quantum computing. He currently serves as an Associate Editor of ACM Transactions on Design Automation of Electronic Systems and ACM Transactions on Embedded Computing Systems. He is an IEEE Fellow, an AAAS Fellow, and an ACM Distinguished Scientist. 
\end{IEEEbiography}

\begin{IEEEbiography}[{\includegraphics[width=1in]{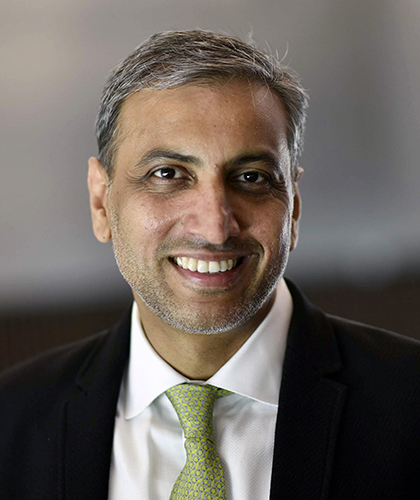}}]{Ankur Srivastava} is the Inaugural Director of Semiconductor Initiatives and Innovation at the University of Maryland. Prior to that he was the seventh director of the Institute for Systems Research (ISR). He has a joint appointment in the Electrical and Computer Engineering Department and ISR. Dr. Srivastava received his B.Tech in Electrical Engineering from Indian Institute of Technology Delhi in 1998 and PhD in Computer Science from UCLA in 2002. He was awarded the prestigious Outstanding Dissertation Award from the CS department of UCLA in 2002. 
\end{IEEEbiography}

\balance

\end{document}